\RequirePackage{fix-cm}
\documentclass[twocolumn]{svjour3}          % twocolumn
\smartqed  % flush right qed marks, e.g. at end of proof
\usepackage[utf8]{inputenc}
\usepackage[english]{babel}
\usepackage{graphicx}
\usepackage{adjustbox}
\usepackage{lipsum} % just for dummy text- not needed for a longtable
\usepackage{longtable}

\usepackage[numbers]{natbib}
\usepackage[ruled, lined, longend, linesnumbered]{algorithm2e}
\usepackage{booktabs, makecell, tabularx}
\usepackage{graphicx}
\usepackage{adjustbox}
\newcommand{\xmark}{\ding{55}}
\usepackage[utf8]{inputenc}
\usepackage[shortlabels]{enumitem}
\usepackage{longtable}
\usepackage{lipsum}
\usepackage{supertabular}
\usepackage{hyperref}
\usepackage{multirow}
\usepackage{soul}
\usepackage{float}
\usepackage{tikz}

\usepackage{amssymb}
\usepackage{pifont}
\usepackage{pgfplots}

\usepackage{wrapfig}

\usepackage{caption}

\usepackage{lipsum}

\begin{document}

\title{ AntibotV: A Multilevel Behaviour-based Framework for Botnets Detection in Vehicular Networks  %\thanks{Grants or other notes
}


\author{Rabah Rahal         \and
        Abdelaziz Amara Korba \and
        Nacira Ghoualmi-Zine \and
        Yacine Challal \and 
        Mohamed Yacine Ghamri-Doudane
}


\institute{Rabah Rahal \at
             Networks and Systems Laboratory (LRS), Badji Mokhtar-Annaba University, Annaba, Algeria \\
              Tel.: +213-674-383-701\\
             \email{rabah.rahal@univ-annaba.org}           %  \\
           \and
          Abdelaziz Amara Korba \at
              Networks and Systems Laboratory (LRS), Badji Mokhtar-Annaba University, Annaba, Algeria \\
              \email{abdelaziz.amara.korba@univ-annaba.org}
              \and
          Nacira Ghoualmi-Zine \at
              Networks and Systems Laboratory (LRS), Badji Mokhtar-Annaba University, Annaba, Algeria \\
              \email{ghoualmi@yahoo.fr}
              \and
          Yacine Challal \at
            Laboratory of Methods of Systems Design (LMCS), Ecole Nationale Supérieure d'Informatique (ESI), Algiers, Algeria \\
            \email{y\_challal@esi.dz}
            \and
          Mohamed Yacine Ghamri-Doudane \at
          Laboratory of Informatics, Image and Interaction (L3i), University of La Rochelle, La Rochelle, France \\
          \email{yacine.ghamri@univ-lr.fr}
}

\date{Received: date / Accepted: date}

\maketitle

\begin{abstract}
Connected cars offer safety and efficiency for both individuals and fleets of private vehicles and public transportation companies. However, equipping vehicles with information and communication technologies raises privacy and security concerns, which significantly threaten the user's data and life. Using bot malware, a hacker may compromise a vehicle and control it remotely, for instance, he can disable breaks or start the engine remotely. In this paper, besides in-vehicle attacks existing in the literature, we consider new zero-day bot malware attacks specific to the vehicular context, WSMP-Flood, and Geo-WSMP Flood. Then, we propose AntibotV, a multilevel behaviour-based framework for vehicular botnets detection in vehicular networks. The proposed framework combines two main modules for attack detection, the first one monitors the vehicle’s activity at the network level, whereas the second one monitors the in-vehicle activity. The two intrusion detection modules have been trained on a historical network and in-vehicle communication using decision tree algorithms. The experimental results showed that the proposed framework outperforms existing solutions, it achieves a detection rate higher than 97\% and a false positive rate lower than 0.14\%.
\keywords{ITS\and Vehicular Networks \and Botnets \and WSMP \and Intrusion detection\and Network Flow\and Controller Area Network \and Machine learning \and Network Forensics\and {Security}}
\end{abstract}

\section{Introduction}
With the proliferation of connected cars, vehicular networks enabled the concept of autonomous fleets of vehicles. Indeed, vehicular networks became a distributed transport fabric capable of making its own decisions about driving customers to their destinations \cite{lee2016internet}. Nowadays, vehicular networks are used in diverses applications ranging from safety such as blind-spot warning to entertainment such as streaming media. Unlike other wireless networks, the nodes in vehicular networks move at high speeds, causing frequent disconnections, and reducing communication time between nodes. In addition, due to mobility, it is possible to move from one environment to another, and each environment has its characteristics (obstacles, signal propagation, etc.) that influence the quality of communication. To adapt to these challenges, specific communication standards have been proposed to ensure efficient communications between vehicles and infrastructure, among these standards we find the dedicated short-range communication standard (DSRC). DSRC is a communication technology that relies heavily on several cooperatives and interoperable standards: IEEE 802.11p, IEEE 1609.x, SAE J2735 Message Set Dictionary, and the emerging SAE J2945.1 standard. IEEE 802.11p and IEEE 1609.4 are used to describe the physical and the Media Access Control (MAC) layer of the system respectively. The network and transport layers are described by IEEE 1609.3, which has two supported stacks (Internet Protocol version 6 (IPv6) for non-safety applications and WAVE Short Message Protocol (WSMP) for safety applications). Security functions and services are described by an IEEE 1609.2 standard protocol. While SAE J2735 and SAE J2945.1 standards are used to define the format of the messages exchanged over the network.

%2. Introduce the specific topic of your research and explain why it is important. 
Vehicular networks can contribute to improving transportation safety and efficiency \cite{Krishnan2010}. However, they raise privacy and security concerns, which significantly threaten the network operations and the user data. One of the most dangerous cybersecurity threats is when the onboard computer of a connected vehicle gets compromised and exploited by a remote attacker. This cyberthreat is known as vehicular bot malware. Unlike the other security threats, it can be used to execute remotely multilevel malicious tasks: (1) distributed network attacks (DDoS) \cite{todorova2016ddos}; (2) violating driver’s data (location privacy \cite{zhou2017security} or illegal GPS tracking, eavesdroping drivers and passengers' conversations \cite{tyagi2014investigating}); (3) controlling the bot vehicles remotely (opening the door, starting the engine, turning on the lights, driving the vehicle away or disabling breaks); (4) misleading the driver by giving false information about the vehicle state (falsifying the fuel level, changing the speedometer reading and displaying failure information on the instrument panel cluster) \cite{ liu2017vehicle}. 

Despite the harmful impact of vehicular bot malwares on the privacy and safety of the driver, there are few researches \cite{garip2016ghost,garip2019shieldnet}, that have considered this issue. Garip et al. \cite{garip2019shieldnet} proposed SHIELDNET, a machine learning-based botnets detection mechanism. As botnet, they considered GHOST \cite{garip2016ghost}, a botnet communication protocol that uses Basic Safety Messages BSMs to dissimulate its communication over the control channel. SHIELDNET detects the use of GHOST, and identifies vehicular botnet communication, by looking for abnormal values of specific BSM fields, which are messages used by security applications only in vehicular networks and cannot be found in other types of networks. Although its effectiveness, SHIELDNET relies on a specific communication protocol (Ghost), thus it would not be effective if the botmaster changes the communication protocol. 

Bot malware is one of the most dangerous cyber threats that can target connected vehicles; a hacker can compromise the on-board computer and take full control of the vehicle. This paper proposes a new intrusion detection system for connected and autonomous vehicles. The proposed system employs a behavioral approach based on machine learning techniques to monitor network and in-vehicle traffic for botnet activity. The detection approach does not suppose a specific botnet communication protocol. Since vehicular bot malware operate within in-vehicle and network communication levels, we propose a multilevel behaviour-based framework for botnets detection in vehicular networks (AntibotV). The proposed framework monitors the interaction of the vehicle with the outside by analyzing the network traffic. It also monitors the in-vehicle activity to detect suspicious operations that may relate to bot malware activity. In addition, this paper present two new bot malware attacks (zero-day attacks) that could be carried out exclusively against vehicular networks communication stack: 1) wave short message protocol flood (WSMP flood); 2) and geographic wave short message protocol-Flood (Geo WSMP flood). Both attacks target the wave short message protocol, which is used in vehicular networks for safety and convenience packets transfer.

The contribution of this paper is two-fold: 
\begin{enumerate}[label=(\roman*)]
\item First, we define and describe a set of zero-day network attacks, information theft scenarios, and in-vehicle threats that can be executed by a hacker through compromising an on-board vehicle computer using bot malwares.
\item Second, we propose a multilevel botnet detection framework based on decision trees that monitors the vehicle's operation at the network and in-vehicle levels to detect the above-mentioned specific attacks as well as more general existing vehicular bot activities.
\end{enumerate}

The rest of the paper is organized as follows. Section \ref{sec:2} provides a background related to vehicular network architecture and botnet communication. Section \ref{sec:rw} summarizes the existing works in the literature related to botnet detection. In section \ref{sec:3} we describe the threat models. Section \ref{sec:4}, describes in detail the proposed framework, AntibotV. In section \ref{sec:5}, we describe the dataset generation steps and discuss the obtained results. Section \ref{sec:6} concludes the paper and draws some line of future work.

\section{Background} \label{sec:2}
This section provides a description of network and in-Vehicle architecture and communication, as well as a brief state of the art about botnet detection. 

\subsection{The architecture of vehicular networks}

A vehicular network organizes and connects vehicles with each other, and with fixed-locations resources (Road Side Units). Many telematics architectures, including navigation, traffic information, entertainment, emergency, and safety services can be provided. In these architectures, traffic information and navigation services are generally provided by central TSPs (Telematics Service Providers) through a vehicle-to-infrastructure communication \cite{chen2010introduction}. On the other hand, the emergency and safety services are supplied by an Onboard Unit (OBU) installed by the individual car manufacturers, to allow mutual communication among different vehicular nodes (vehicle-to-vehicle) \cite{zhou2015secure}.

\subsubsection{Network stack architecture}

\begin{figure*}[h!]
\begin{adjustbox}{width=0.75\textwidth, center}
   \includegraphics{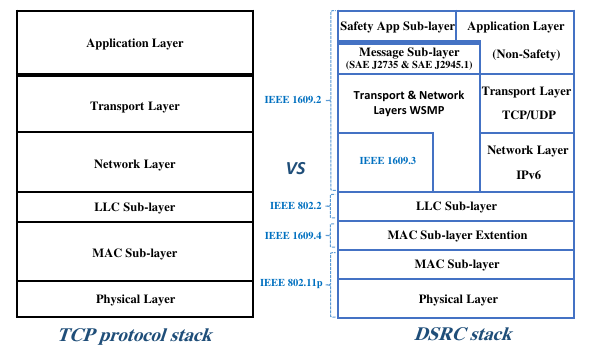}
\end{adjustbox}
    \caption{TCP vs DSRC stack}
   \label{fig:tcpvsdsrc}
\end{figure*}

\begin{figure}[h!]
    \centering
    \includegraphics[width=0.4\textwidth]{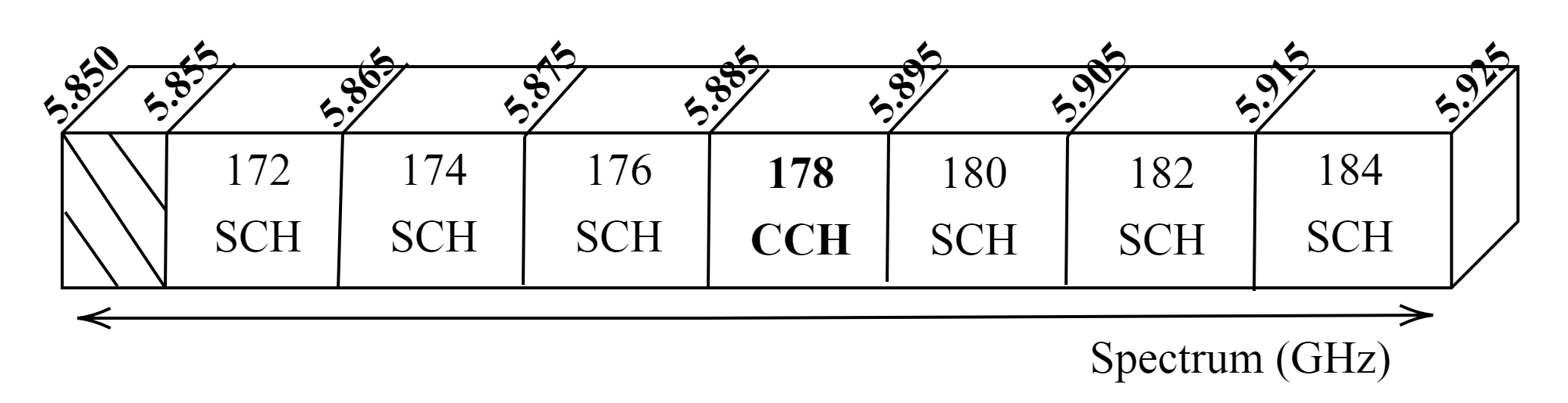}
    \caption{WAVE radio channels}
    \label{fig:waveradiochannels}
\end{figure}

To ensure vehicle-to-vehicle and vehicle-to-infra-structure communication, the automotive industry has developed the dedicated short-range communication standard (DSRC) (Figure \ref{fig:tcpvsdsrc}). A communication technology that relies heavily on several cooperatives and interoperable standards \cite{kenney2011dedicated}. At the PHY and MAC layers, DSRC utilizes IEEE 802.11p Wireless Access for Vehicular Environments (WAVE).  WAVE is an approved amendment of the IEEE 802.11 standard that enables secure wireless communication and physical access for high speed (up to 27 MB/s), short-range (up to 1000 m), and low latency. The spectrum allocated to it is from 5.850 to 5.925 GHz, divided into seven 10 MHz channels \cite{ahmed2013overview}. Channel 178 is reserved for control information and the others six channels are used for service applications (Figure \ref{fig:waveradiochannels}).

At the MAC sublayer extension of DSRC, the IEEE 1609.4 standard is deployed. It is used for priority access, management services, channel switching, and routing \cite{ieee1609.4} in order to enable to operating efficiently on seven channels alternately \cite{song2017performance}. As regards the IEEE 1609.2 standard, it includes techniques used to secure application messages and describes administrative functions necessary to support the core security functions \cite{ieee1609.2}.

At the Network Layer, DSRC uses the IEEE 1609.3 standard. A standard that supports two protocol stacks, Wave Short Message Protocol (WSMP) and IPv6. The choice between using WSMP or IPv6+UDP/TCP depends on the requirements of the application. For the ones that depend on the routing capabilities to transmit multi-hop packets like commercial applications, IPv6, UDP, and TCP are used. However, the applications that require the transfer of single-hop messages like security and convenience applications, WSMP protocol is used. 
Unlike the IP, UDP, and TCP protocols, the WSMP is a WAVE network layer unique protocol and can be found only in vehicular networks. It is used only to support high-priority and time-sensitive communication. 

Finally, for the format of the messages exchanged over DSRC, like data frames and elements used by the applications, they are defined in the SAE J2735 and SAE J2945.1 standards. SAE J2735 represents a dataset that contains message syntax. It contains many types, among others, we mention the Basic Safety Message (BSM) (periodically transmitted to provide current information and status) and the Common Safety Request (CSR) \cite{hedges2008overview}. Other messages norms for the V2V safety applications are specified with the SAE J2945.1 standard.

%%%%%%%%%%%%%%%%%%%%%%%%%%%%%%%%%%%%%%%%%%%%%%%%%%%%%%%%%%%%%%%%%
%%%%%%%%%%%%%%%%%%% Test %%%%%%%%%%%%%%%%%%%%%%%%%%%%%%%%%%%%%%%%
\subsubsection{In-Vehicle architecture}
\begin{figure*}[h!]
    \centering
    \includegraphics[width=0.7\textwidth]{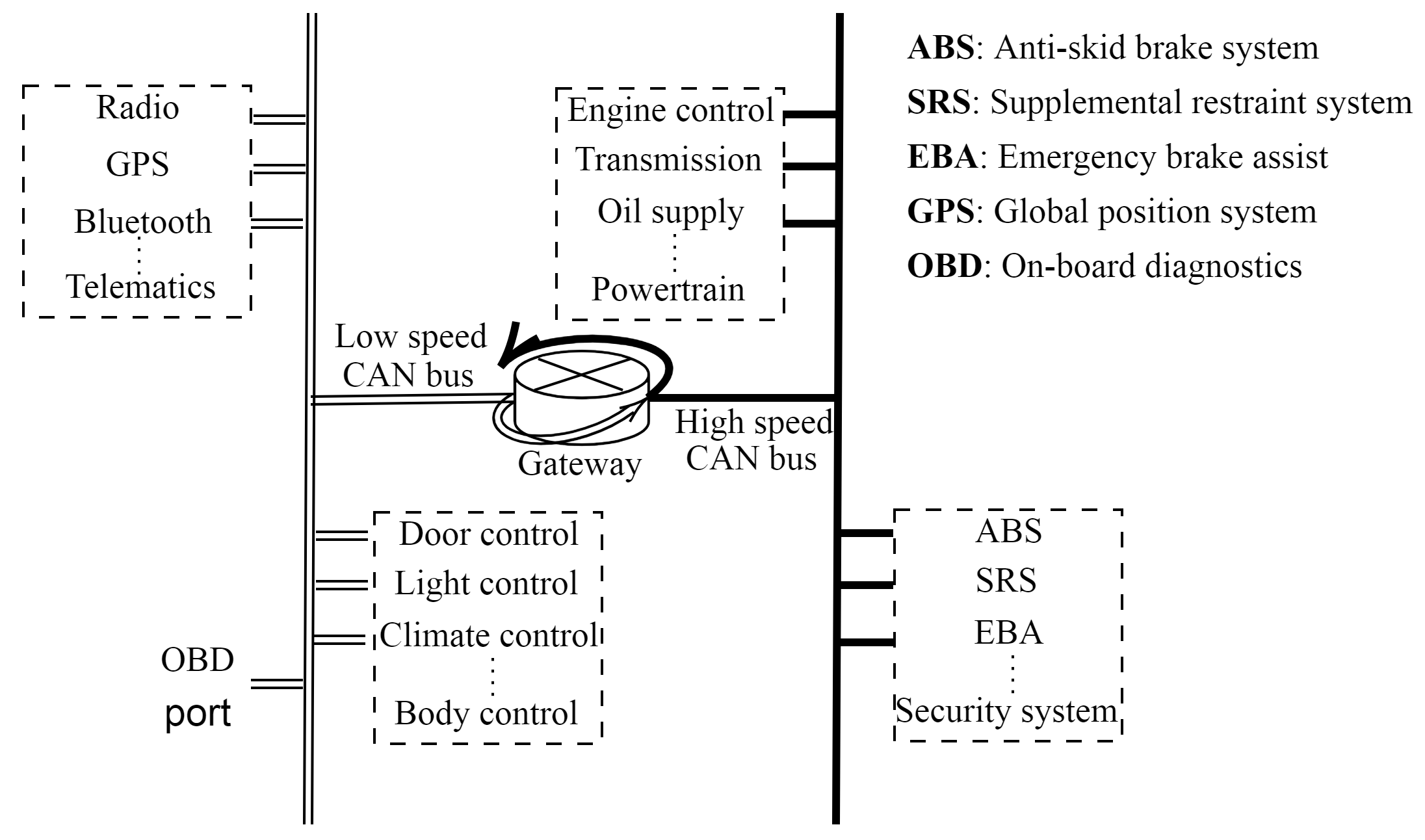}
    \caption{In-vehicle network}
    \label{fig:invehiclenetwork}
\end{figure*}

Nowadays, we are witnessing the automotive industry converging to replace the mechanical components of the vehicle with other electronic components labeled electronic control units (ECU), the number of deployed units is expected to reach 3.29 billion by 2025 \cite{ECUStat}. ECUs simplify the interior architecture of vehicles, thus repair and diagnose even for those who know nothing about vehicles. Each ECU contains its sensors and actuators, it receives input from its sensors and implements specific functions by its actuators. Communication between these ECUs is ensured through a dedicated bus type especially for vehicular networks; called Controller Area Network (CAN bus). ECUs and CAN buses together form the In-vehicle network (Figure \ref{fig:invehiclenetwork}). 

In this In-vehicle network, there are two types of CAN bus (high-speed and low-speed) connected by a gateway \cite{liu2017vehicle}. For the communication of critical modules (power train, brake, etc.), the high-speed CAN bus is used. For the other types of modules (telematics, body control, etc.) a low-speed CAN bus is used. The transmission on CAN bus is done sequentially. However, if more than one device transmits at the same time, a media access control (MAC) protocol Carrier-sense multiple access with Collision Resolution (CSMA-CR) is used. CSMA/CR uses priorities in the frame header to avoid collisions \cite{can_intro}.

\subsection{Botnets} 

A botnet is a collection of internet-connected devices (computers, smartphones, IP cameras, routers, IoT equipment, etc.) called bots that are infected by a malware, to be controlled remotely by an operator (botmaster) without users' knowledge. The communication between the botmaster and the bots is ensured through the Command and Control (C\&C) server. Botnets can be used to achieve harmful attacks, such as: launching Distributed Denial-of-Service (DDoS) on rival websites or services, send spam, distribute malwares, stole user/equipment private information, and applying interior activities on the infected devices. For example, the Mirai botnet \cite{f} in 2016, was able to carry out a massive DDoS attack that brought down major sites like Amazon, Netflix, Paypal, and Reddit \cite{ops}.

\section{Related Work} \label{sec:rw}
Researchers have worked on detecting botnets and overcoming their negative impact. They proposed methods suited to the characteristics of each type of network (communication stack, protocols, characteristics of equipment, etc). Other worked on the protection against DDoS attacks that can be caused by a botnet \cite{correa2021ml, agrawal2021sdn, alhisnawi2020detecting, rahal2020towards}. We find also different anomaly detection, revolutionary, and hybrid classification techniques proposed to deal with new types of botnets \cite{otoum2021ids, oseni2021security, liu2020privacy, dibaei2020investigating, Zhuo2017, Liu2015, Lakhina2004, hu2020cablemon, Li2006}.

In \cite{correa2021ml}, the authors proposed an ML-based DDoS detection and identification approach using native cloud telemetry macroscopic monitoring. A lightweight method and completely agnostic to specific protocols and services, which can detect any kind of DDoS attack that target the resources without the need for previous training. The authors in \cite{agrawal2021sdn} have worked on the detection of Low-Rate DDoS (LR-DDoS) attack (exactly the Shrew attack). They proposed a new mechanism which not only detects and mitigates the shrew attack but traces back the location of the attack sources as well. The attack is detected using the information entropy variations, and the attack sources are traced back using the deterministic packet marking scheme. If the DDoS attack is caused by a botnet, the traceability mechanism can be used to identify bot nodes in the network. Approaches that deal with DDoS can be used to mitigate the effects of botnets (and even to identify bot nodes), however, if the botnet is used for other attacks (e.g. information theft), DDoS detection techniques will not be effective.

In \cite{otoum2021ids}, the authors proposed a model (known as “AS-IDS”) that combines two detection approaches (anomaly-based and signature-based) to detect known and unknown attacks in IoT networks. The proposed model has three phases: traffic filtering, preprocessing, and hybrid IDS. The signature-based IDS subsystem investigates packets by matching the signatures and categorized them as an intruder, normal or unknown. The anomaly-based IDS subsystem employs Deep Q-learning to identify unknown attacks. In \cite{biswas2021botnet}, The proposed method attempts to identify malicious botnet activity from legitimate traffic using novel deep learning approaches, exactly Gatted Recurrent Units (GRU) model. In a second step, they introduced adjusted hyperparameters to make the computational complexity low and more accurate than the existing models. However, the use of two detection approaches could cause an unbearable overload, which does not suit the type of network whose nodes have limited computing and storage capacities.

In the case of traditional computer-based botnet such as Ebury botnet \cite{f}, network-based detection is one of the most effective detection techniques that has attracted the attention of many researchers \cite{sinha2019tracking, zhao2013botnet, strayer2008botnet,ranjan2014machine, ranjan2013machine}. By analyzing the rate of failed connection and network flow features, the network-based detection technique identifies the exchanged traffic between the C\&C server and the bots.

Regarding smartphone-based botnet (like WireX \cite{f}), several solutions have been proposed to tackle their impact, including rule-based \cite{ongtang2012semantically} and signature-based \cite{zhao2012robotdroid} approaches. However, it is the behavioural analysis methods that have caught the attention of researchers \cite{andronio2015heldroid, enck2014taintdroid, ni2020security}. At different levels of the Android OS stack, different sets of behavioural data are available. By monitoring several layers of the Android OS, the behaviour-based detection method seeks to detect malicious applications (API calls, etc.).

Concerning IoT-based botnets, several methods have been proposed to fight against them by taking into account their special characteristics (communication architecture, protocols, etc.) \cite{iou2018signature, dwyer2019profiling, li2019analysis, ridley2019machine, wazid2020lam, jangirala2020designing, bera2021designing}. In \cite{iou2018signature}, the authors proposed a lightweight signature-based method that aims to ensure a reduced false positive rate without adding an overhead to IoT devices. Other researchers have worked on network-level detection (exactly, the DNS protocol) to ensure general solutions that cannot be affected by any hardware specificity \cite{dwyer2019profiling,li2019analysis,ridley2019machine, wazid2020lam, jangirala2020designing}.

The existing botnets detection mechanisms cannot be applied directly for vehicular networks due to differences in terms of communication stack, protocols, frames format, and architecture. In the context of vehicular networks, to the best of our knowledge, there is only one research \cite{garip2016ghost} that has tackled the problem of botnets detection. Garip et al. \cite{garip2016ghost} focused on the communication protocol between the botmaster and the bot vehicles. They investigated the usage of periodic basic safety messages to transmit commands from the botmaster to the bot vehicles. This communication protocol called "Ghost" allows the hacker to hide its remote communication with the infected vehicle. The authors considered two bot malware attacks feasible against vehicular networks: Botveillance \cite{garip2018botveillance}, and congestion attacks \cite{garip2015congestion}. Botveillance is an adaptive cooperative surveillance attack against pseudonymous systems in vehicular networks, which is based on vehicular botnet and performed by vehicles themselves without depending on any additional hardware. It is used to track vehicles of interest or violate the privacy of drivers. On the other hand, the congestion attack is based on the vehicular congestion avoidance application. It uses bot vehicles to spread wrong congestion information to the other legitimate vehicles, to cause congestion on specific roads or areas.

The same authors \cite{garip2019shieldnet} proposed SHIELDNET, a vehicular botnets detection mechanism that uses machine learning techniques to detect GHOST usage and identify the botnet communication. SHIELDNET detects botnet activity by looking for anomalous values of specific BSM fields. Although the efficiency of the proposed solution, the Ghost protocol is a single-hop range, which makes the command of the botmaster sent only to its neighbours, making the range of communication very short and unreachable by far bot vehicles. Moreover, if the botnet changes the communication protocol, it will not be effective. Furthermore, the scenario of the botnet in vehicular networks is not real. Because transmitting commands through V2V communication will make the botmaster control capacity very limited and it needs to be on the road and near other botnet nodes. Therefore, the ideal way to imagine a botnet in vehicular networks is where the botmaster is anywhere and can control those vehicles using V2I communication and send commands through infrastructure. On the other hand, if the installed malware is used to apply in-vehicular activities, SHIELDNET will not be able to detect any abnormal behaviour.

To the best of our knowledge, no study of botnets in vehicular networks has combined bot malwares activities at the network and in-vehicle levels. Furthermore, existing solutions in the literature are based on specialized protocols, limiting them to certain cases and making them non-general. In this paper, we propose AntibotV, a multilevel behaviour-based framework for botnets detection in vehicular networks that monitors the vehicle's interaction with the outside by analyzing the network traffic. It also monitors the in-vehicle activity to detect suspicious operations that may relate to bot malware activity. Furthermore, AntibotV does not suppose a particular botnet communication protocol, making it a general botnet-detection model regardless of the botnet communication protocols.

\section{Threat Models} \label{sec:3}

\begin{figure*}[h!]
\begin{adjustbox}{width=1\textwidth, center}
    \includegraphics{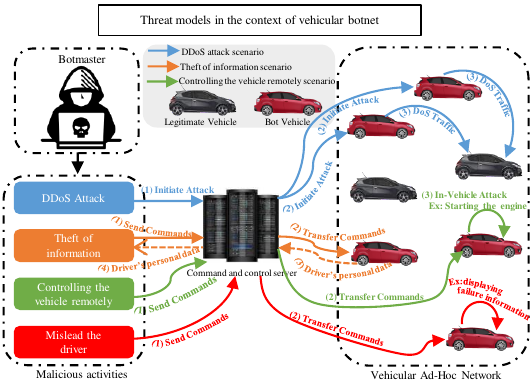}
\end{adjustbox}
    \caption{Vehicular Botnets}
    \label{fig:vb}
\end{figure*}

In this section, we provide a detailed description of the three categories of cyberattacks that can be executed against a target vehicle using bot malwares. First, we provide a detailed description of two zero-day DDoS attacks. Then based on existing attacks on privacy, we define new scenarios applicable to the vehicular context. Finally, we present some in-vehicle attacks that exist in the literature. Figure \ref{fig:vb} shows the different categories of cyberattacks that a hacker may execute using bot malware.

\subsection{DDoS attacks}
Due to the differences between DSRC stack and the TCP/IP stack, it is important to consider DDoS attack scenarios specific to the vehicle network, and not be limited to DDoS attacks common to all IP networks. Therefore, in this paper, we consider two zero-day attacks specific to vehicle networks. Both attacks exploit the WAVE Short Message Protocol (WSMP) which is used by security and traffic management applications for the transfer of critical data such as vehicle speed, kinematic state, etc. Both attacks can prevent the transfer of safety messages between vehicles and thus cause catastrophic damage.

\begin{figure}[h!]
    \centering
    \includegraphics[width=0.4\textwidth]{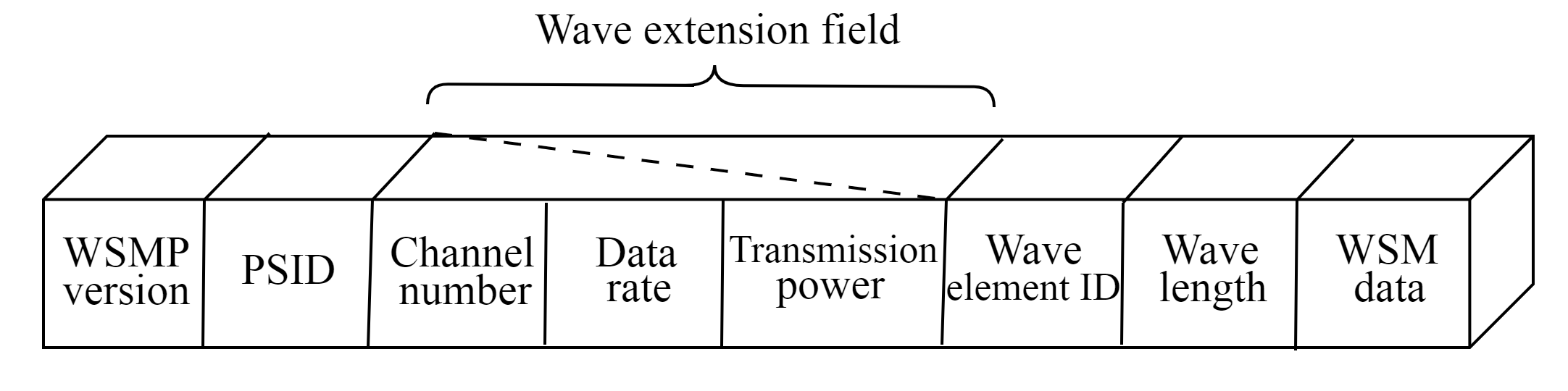}
    \caption{WSM frame}
    \label{fig:wsmframe}
\end{figure}

The WSM packets (Figure \ref{fig:wsmframe}) exchanged between vehicles are composed of the following fields: 
WSMP version that gives the version of the protocol, channel number, and data rate to specify which channel and data rate are used for the transmission, WAVE element ID represents the WSMP header, WAVE Length to specify the length of the packet, and the WSM Data field contains the payload data. The Provider Service Identifier field (PSID), identifies the service that the WSM payload is associated with. For example, if an application tries to get access to the WAVE service, it should be registered with its unique Provider Service Identifier (PSID). The WAVE provider devices use PSID in their announcement messages to indicate that a certain application is provided by this device. On the other hands, as the vehicle passing the roadside device, user devices, which may host such application, upon reception of such announcement, compares to check if there is a match between the PSIDs in the announcements and PSIDs in its tables, then the vehicle establishes communication with that roadside unit \cite{kenney2011dedicated, ahmed2013overview}.

In both attacks, the hacker attempts to misuse the WSMP protocol. In the first attack, WSMP flood, the hacker sends to the victim vehicle WSM packets with unknown PSID field values (not associated with any WAVE service). Upon the reception of the forged WSM packet, the target vehicle attempts to check (lookup) for the corresponding entry within its PSID/Service table. Checking the PSID of one WSMP frame will not be a problem. However, checking the PSID of a huge amount of WSM packets at the same time will exhaust the resources of the target vehicle and make it unable to respond or receive legitimate security and convenience application packets.

\begin{figure*}[h!]
\begin{adjustbox}{width=0.9\textwidth, center}
    \includegraphics{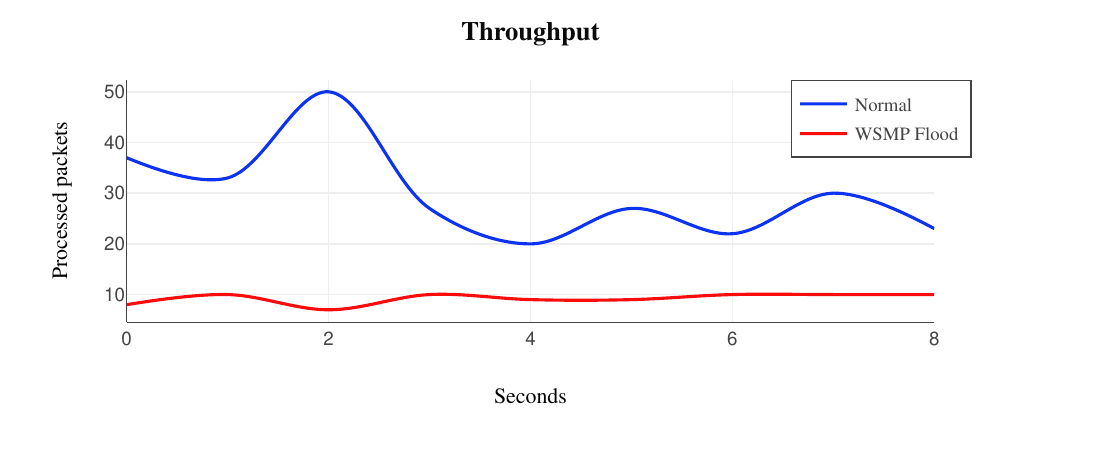}
\end{adjustbox}
    \caption{Normal throughput vs WSMP Flood throughput}
    \label{fig:thWsmpF}
\end{figure*}

The second attack, geographic WSMP flood, has the same operating mode as WSMP flood, but wider impact. The hacker broadcasts the forged WSMP messages to all its neighbours within a specific geographic area. Thus, it exhausts the resources of several vehicles, and it consumes the bandwidth of an entire geographic area. Before validating AntibotV with the WSMP flood attack, we tested it, and we found that it can decrease the throughput by 63\% (the number of security and convenience frames arrived), which makes the attack very effective (as shown in figure \ref{fig:thWsmpF}). More details about the experiment setting (scenario, number of nodes, time, etc.) are provided in section \ref{sec:5.1.1}. 

\begin{table*}[!htbp]
  \caption{Threat models}
  \label{tab:threatmod}
\begin{adjustbox}{width=0.9\textwidth, center}
\begin{tabular}{|l|l|l|}
\hline
Category                                                                        & Attack             & Objective                                                                                                                                                                                       \\ \hline
\multirow{2}{*}{DDoS}                                                           & WSMP Flood         & \begin{tabular}[c]{@{}l@{}}Exhausting the resources of the target vehicle and make it unable \\ to respond or receive legitimate security and convenience \\ applications packets.\end{tabular} \\ \cline{2-3} 
                                                                                & Geo-WSMP Flood     & \begin{tabular}[c]{@{}l@{}}Exhausting the resources of several vehicles, and consuming\\ the bandwidth of an entire geographic area.\end{tabular}                                               \\ \hline
\multirow{2}{*}{\begin{tabular}[c]{@{}l@{}}Theft of\\ information\end{tabular}} & GPS Tracking       & Tracking the vehicle’s location on real time.                                                                                                                                                   \\ \cline{2-3} 
                                                                                & Audio Surveillance & \begin{tabular}[c]{@{}l@{}}Recording the conversations using the vehicle’s internal microphones, \\ and sends them to the botmaster.\end{tabular}                                               \\ \hline
\multirow{3}{*}{In-vehicle}                                                     & Frame Falsifying   & \begin{tabular}[c]{@{}l@{}}Fabricate fake frames that contain erroneous data to mislead the driver: \\ falsifying the fuel level, changing the speedometer reading…\end{tabular}                \\ \cline{2-3} 
                                                                                & Replay Attack      & \begin{tabular}[c]{@{}l@{}}Replaying captured frames. Could be used to: opening the door, starting \\ the engine…\end{tabular}                                                                  \\ \cline{2-3} 
                                                                                & DoS Attack         & \begin{tabular}[c]{@{}l@{}}Monopolize the transmission channel to delay or prevent the transmission\\ of legitimate CAN frames. Example: disable the brakes.\end{tabular}                        \\ \hline
\end{tabular}
\end{adjustbox}
\end{table*}

\subsection{Theft of information}
A team of academics from the University of California at San Diego and the University of Washington has found out that the audio surveillance inside the vehicles is possible \cite{checkoway2011comprehensive}. In such situations, the attacker indirectly uses virtual assistance tools (like Siri \cite{apple}) by giving it malicious commands hidden in recorded music, innocuous-sounding speech \cite{siri_hiddencmd} or a low-powered laser \cite{sugawara2019light}. Siri begins recording the conversations using the vehicle's internal microphones and sends them to a remote server belonging to the hacker every 10 or 20 seconds\cite{murphy_2017}. 

In the case of GPS tracking, the hacker may attempt to track the vehicle's location in real-time, check the location, or retrieve the complete trajectory. For real-time tracking, the GPS information must be sent every second (streaming) to the botmaster. For the other two types of GPS tracking, the information are sent periodically or on-demand. In this paper, we consider the real-time GPS tracking scenario. The bot vehicle sends the latitude (4 bytes), longitude (4 bytes), speed (2 bytes), time (2 bytes) and direction coordination in streaming to the botmaster \cite{thomas2020c}. 

\subsection{In-Vehicle attacks}
The in-vehicle networks brought convenience to manufacturers, drivers, and after-sales services. However, they raise vulnerabilities \cite{liu2017vehicle} that can be exploited by a hacker to conduct the following attacks:
\begin{enumerate} 
   \item \textbf{Frame Falsifying and injection}: CAN frames are sent in plaintext, which allows the bot malware within the vehicle to retrieve and analyze their contents. Using the captured data, the bot malware can fabricate fake frames that contain erroneous data to mislead the ECUs. Subsequently, the fabricated frames are injected in CAN bus to carry out malicious activities such as: falsifying the fuel level, changing the speedometer reading, or displaying failure information that may mislead the driver.

   \item \textbf{Replay Attack}: The frames are transmitted through CAN bus using broadcast and without authentication. Thus, CAN frames can be easily captured by the bot malware to be analyzed and replayed in a second step.  In \cite{koscher2010experimental}, the authors found that the range of valid CAN frames is small. Hence, by iteratively testing CAN frames (i.e., the fuzzing test mentioned in \cite{koscher2010experimental}), adversaries can discover many functions of selected ECUs. The replay attack could be used to open the door, start the engine, turn on the lights or remotely drive the vehicle.

    \item \textbf{DoS Attack}: messages transmitted with the smallest identifier are the messages with the highest priority. The bot malware could use this vulnerability to monopolize the transmission channel, and thus delay or prevent the transmission of legitimate CAN frames. For instance, the bot malware can prevent a CAN frame from being transmitted to the ECU of brakes, which could lead to accident \cite{liu2017vehicle}.
    
\end{enumerate}

Table \ref{tab:threatmod} provides a summary of the different security threats discussed earlier.

\section{AntibotV Framework}\label{sec:4}

In this section, we describe in detail the AntibotV, a multi-level framework to detect vehicular botnets. Moreover, we describe the detection process based on machine-learning algorithms.

\subsection{AntibotV overview}

We suppose that network traffic generated by a bot malware is different from legitimate traffic. In other words, the bot malware alters the traffic pattern of the vehicle. This hypothesis comes from the fact that connected cars run specialized applications related to safety and convenience, such as cooperative collision warning, V2V post-crash notification, congested road notification, etc... Running specialized and particular applications, makes connected vehicles' network traffic pattern regular as long as it is not compromised. Thus, we believe bot malware that compromises a connected vehicle should alter its network traffic pattern. Likewise, at the in-Vehicle level, because of the regularity of communication patterns, it is possible to identify a legitimate in-vehicle traffic pattern, and thus, detect malicious bot activity.

\begin{table*}[!htbp]
\small
  \caption{List of network features}
  \label{tab:features_network}
\begin{adjustbox}{width=1\textwidth, center}
\begin{tabular}{|l|l|l|}
\hline
\multicolumn{2}{|l|}{{\color[HTML]{00000A} \textbf{Features}}}                                    & {\color[HTML]{00000A} }                                                                                    \\ \cline{1-2}
{\color[HTML]{00000A} \textbf{Nb}} & {\color[HTML]{00000A} \textbf{Name}}                         & \multirow{-2}{*}{{\color[HTML]{00000A} \textbf{Description}}}                                              \\ \hline
{\color[HTML]{00000A} 1}           & {\color[HTML]{00000A} Flow Duration}                         & {\color[HTML]{00000A} Duration of the flow in microsecond}                                                 \\ \hline
{\color[HTML]{00000A} 2}           & {\color[HTML]{00000A} Flow IAT Mean}                         & {\color[HTML]{00000A} Mean time between two packets sent in the flow}                                      \\ \hline
{\color[HTML]{00000A} 3}           & {\color[HTML]{00000A} Flow IAT Std}                          & {\color[HTML]{00000A} Standard deviation time between two packets sent in the flow}                        \\ \hline
{\color[HTML]{00000A} 4}           & {\color[HTML]{00000A} Flow IAT Max}                          & {\color[HTML]{00000A} Maximum time between two packets sent in the flow}                                   \\ \hline
{\color[HTML]{00000A} 5}           & {\color[HTML]{00000A} Flow IAT Min}                          & {\color[HTML]{00000A} Minimum time between two packets sent in the flow}                                   \\ \hline
{\color[HTML]{00000A} 6}           & {\color[HTML]{00000A} Fwd IAT Tot}                           & {\color[HTML]{00000A} Total time between two packets sent in the forward direction}                        \\ \hline
{\color[HTML]{00000A} 7}           & {\color[HTML]{00000A} Fwd IAT Mean}                          & {\color[HTML]{00000A} Mean time between two packets sent in the forward direction}                         \\ \hline
{\color[HTML]{00000A} 8}           & {\color[HTML]{00000A} Fwd IAT Std}                           & {\color[HTML]{00000A} Standard deviation time between two packets sent in the forward direction}           \\ \hline
{\color[HTML]{00000A} 9}           & {\color[HTML]{00000A} Fwd IAT Max}                           & {\color[HTML]{00000A} Maximum time between two packets sent in the forward direction}                      \\ \hline
{\color[HTML]{00000A} 10}          & {\color[HTML]{00000A} Fwd IAT Min}                           & {\color[HTML]{00000A} Minimum time between two packets sent in the forward direction}                      \\ \hline
{\color[HTML]{00000A} 11}          & {\color[HTML]{00000A} Down/Up Ratio}                         & {\color[HTML]{00000A} Download and upload ratio}                                                           \\ \hline
{\color[HTML]{00000A} 12}          & {\color[HTML]{00000A} Idle Mean}                             & {\color[HTML]{00000A} Mean time a flow was idle before becoming active}                                    \\ \hline
{\color[HTML]{00000A} 13}          & {\color[HTML]{00000A} Idle Std}                              & {\color[HTML]{00000A} Standard deviation time a flow was idle before becoming active}                      \\ \hline
{\color[HTML]{00000A} 14}          & {\color[HTML]{00000A} Idle Min}                              & {\color[HTML]{00000A} Minimum time a flow was idle before becoming active}                                 \\ \hline
{\color[HTML]{00000A} 15}          & {\color[HTML]{00000A} Tot Fwd Pkts}                          & {\color[HTML]{00000A} Total packets in the forward direction}                                              \\ \hline
{\color[HTML]{00000A} 16}          & {\color[HTML]{00000A} Flow Pkts/s}                           & {\color[HTML]{00000A} Number of flow bytes per second}                                                     \\ \hline
{\color[HTML]{00000A} 17}          & {\color[HTML]{00000A} Fwd Pkts/s}                            & {\color[HTML]{00000A} Number of forward packets per second}                                                \\ \hline
{\color[HTML]{00000A} 18}          & {\color[HTML]{00000A} Bwd Pkts/s}                            & {\color[HTML]{00000A} Number of backward packets per second}                                               \\ \hline
{\color[HTML]{00000A} 19}          & {\color[HTML]{00000A} total\_Bwd\_Packets}                   & {\color[HTML]{00000A} The total number of packet in the bwd direction.}                                    \\ \hline
{\color[HTML]{00000A} 20}          & {\color[HTML]{00000A} total/min/max/mean/std\_f/b\_Pktl}     & {\color[HTML]{00000A} The size of packets and the standard deviation size in fwd or bwd direction.}        \\ \hline
{\color[HTML]{00000A} 21}          & {\color[HTML]{00000A} Avg\_Packet\_Size}                     & {\color[HTML]{00000A} The mean packets size.}                                                              \\ \hline
{\color[HTML]{00000A} 22}          & {\color[HTML]{00000A} f/b\_AvgSegmentSize}                   & {\color[HTML]{00000A} The median noticed size in the fwd or bwd direction.}                                \\ \hline
{\color[HTML]{00000A} 23}          & {\color[HTML]{00000A} f/b\_AvgPacketsPerBulk}                & {\color[HTML]{00000A} The mean number of packets bulk rate in the fwd or bwd direction.}                   \\ \hline
{\color[HTML]{00000A} 24}          & {\color[HTML]{00000A} f/b\_AvgBulkRate}                      & {\color[HTML]{00000A} The average number of bulk rate in the fwd or bwd direction}                         \\ \hline
{\color[HTML]{00000A} 25}          & {\color[HTML]{00000A} act\_data\_pkt\_forward}               & {\color[HTML]{00000A} The number of packets with at least 1 byte of TCP data payload in the fwd direction} \\ \hline
{\color[HTML]{00000A} 26}          & {\color[HTML]{00000A} min\_seg\_size\_forward}               & {\color[HTML]{00000A} The smallest segment size noticed in the fwd direction.}                             \\ \hline
{\color[HTML]{00000A} 27}          & {\color[HTML]{00000A} Total\_f/b\_headers}                   & {\color[HTML]{00000A} The total number of bytes used for headers in fwd or bwd direction}                  \\ \hline
{\color[HTML]{00000A} 28}          & {\color[HTML]{00000A} f/b\_AvgBytesPerBulk}                  & {\color[HTML]{00000A} The mean number of bytes bulk rate in the fwd or bwd direction.}                     \\ \hline
{\color[HTML]{00000A} 29}          & {\color[HTML]{00000A} Init\_Win\_bytes\_forward/backward}    & {\color[HTML]{00000A} The total number of bytes sent in initial window in the fwd or bwd direction.}       \\ \hline
{\color[HTML]{00000A} 30}          & {\color[HTML]{00000A} total/min/max/mean/std\_Bwd\_iat}      & {\color[HTML]{00000A} The total, max, min, mean, and standard time between packets for the bwd direction.} \\ \hline
{\color[HTML]{00000A} 31}          & {\color[HTML]{00000A} f/b\_psh/urg\_cnt}                     & {\color[HTML]{00000A} The number of the PSH or URG flags were set in packets in the fwd or bwd direction.} \\ \hline
{\color[HTML]{00000A} 32}          & {\color[HTML]{00000A} Min/mean/max/std\_active}              & {\color[HTML]{00000A} The min, max, mean, and std time a flow was active before becoming idle.}            \\ \hline
{\color[HTML]{00000A} 33}          & {\color[HTML]{00000A} Flow\_Byts\_PerSecond}                 & {\color[HTML]{00000A} The number of a flow bytes per second.}                                              \\ \hline
{\color[HTML]{00000A} 34}          & {\color[HTML]{00000A} min/max\_flowpktl}                     & {\color[HTML]{00000A} The length (min, max) of flow.}                                                      \\ \hline
{\color[HTML]{00000A} 35}          & {\color[HTML]{00000A} Flow\_fin/syn/rst/psh/ack/urg/cwr/ece} & {\color[HTML]{00000A} The number of packets with flags.}                                                   \\ \hline
{\color[HTML]{00000A} 36}          & {\color[HTML]{00000A} Sflow\_f/b\_Packet}                    & {\color[HTML]{00000A} The average number of packets in a sub flow for the fwd or bwd direction.}           \\ \hline
{\color[HTML]{00000A} 37}          & {\color[HTML]{00000A} Sflow\_f/b\_Bytes}                     & {\color[HTML]{00000A} The average number of bytes in a sub flow for the fwd or bwd direction.}           \\ \hline
\end{tabular}
\end{adjustbox}
\end{table*}

% 2) == Valice is HIDS: advantages of HIDS
The AntibotV is a host-based intrusion detection system, running the system in a vehicle, rather than at some point of the network. The proposed system monitors communication with outside through analyzing network traffic, and it monitors in-vehicle communication by analyzing CAN bus frames. The two-level monitoring allows an effective detection of bot malware activity which consists of sending data and receiving commands from the botmaster, at the network level, and executing control commands, falsify the fuel level, change the speedometer reading or
display failure information at the in-vehicle level.

% 3) == Characterization of network traffic using a set of network features 
The network traffic is characterized using a set of statistical features such as flow duration, total packets sent in the forward direction, minimum packet length, etc (a detailed description of used features is provided in the next section). To characterize the in-vehicle communication a set of twelve features is used including timestamp (recorded time), CAN ID (identifier of CAN message in HEX), DLC (number of data bytes, from 0 to 8), DATA[0~7] (data value in byte).

%model générél applicable dans tous les cas
Regarding network traffic, it is independent of the type of vehicle, model, or year and it depends only on the protocol stack used, which implies that AntibotV is applicable in all vehicles that use DSRC standard as a wireless access technology. Concerning CAN bus standard, it consists of two versions based on the length of the arbitration field. CAN bus 2.0A defines an 11-bit standard frame format while CAN bus 2.0B is compatible with data messages in a standard frame and extended frame format. Our framework mainly focuses on CAN ID, DLC, and the DATA fields (Table [\ref{tab:canbusattrib}]), and the three fields are found in both versions (2.0A and 2.0B) of CAN bus protocol and they have the same length. This implies that AntibotV is a general framework for all vehicles that use the standard DSRC as wireless access technology and CAN bus protocol with its two versions, regardless of the model of the vehicle.

% 4) == Model induction and continious monitoring of the vehicle behaviour 
Being a host-based system, AntibotV should not consume too many of the vehicle’s resources, because that would impact its other operations. For that reason, we move the computational load of the training classification model to a centralised server on the cloud. Thus, the vehicle runs an already trained classification model, which would minimize the resource consumption of the vehicle. Additionally, unlike signature-based models, AntibotV is a behaviour-based model based on machine learning algorithms, which means there is no overhead to maintain a signature database up to date, nor huge computational resources like deep learning-based models. In addition, it does not require any hardware modification.

% 5) === Architecture ====
AntibotV has a modular architecture inspired by the IDMEF (Intrusion Detection Message Exchange Format) architecture proposed by the IDWG group \cite{wood2002intrusion}. The architecture is mainly composed of three modules: traffic collection module, analyzer module, and manager module. The first module collects network traffic and in-Vehicle CAN frames. The analyzer module is responsible for analysing the vehicle’s traffic data. The manager module handles alerts, sends notifications, and updates the classification model for both traffic. A detailed description of these modules is provided in the next sections. 

\subsection{Traffic collection module}
Used to collect and process the vehicle traffic and apply several pre-treatment operations to extract from the raw data a vector of features, which will be used by the analyzer module. It collects two types of traffic: network flow and CAN bus frames.

\subsubsection{Network traffic collector}

This module collects information about network traffic from exchanged packets. Each packet is mapped to a network flow identified by five attributes namely source IP, destination IP, source port, destination port, and protocol. The RFC 3697 \cite{ipv4} defines traffic flow as "a sequence of packets sent from a particular source to a particular unicast, anycast, or multicast destination that the source desires to label as a flow”. TCP flows are usually terminated upon connection teardown (by FIN packet) while UDP flows are terminated by a flow timeout. Table \ref{tab:features_network} shows the list of network flow-based features extracted for each network flow. In the end, the network traffic collector generates a vector of calculated features, then transfers it to the analyzer module.

\subsubsection{In-vehicle traffic collector}

The in-vehicle traffic collector module analyses frames exchanged on CAN bus. Unlike the network traffic, which is processed as flows, the in-vehicle traffic is analysed using the deep frames inspection technique. The analysis is carried out through real-time observation of frames as they traverse CAN bus links. A CAN bus frame contains the following fields: the Start of Frame (1b), Message-ID (an 11b identifier that represents the message priority), Control fields (3b), Data Length (number of data bytes, 4b, from 0 to 8), Data[0~7] (Data to be transmitted, 0-64b), CRC (15b), ACK fields (3b), End of Frame Delimiter (7b) \cite{can:link}.  

The in-vehicle traffic collector generates a features vector with the following fields: timestamp (recorded time), CAN ID, DLC, and the DATA, as illustrated in table \ref{tab:canbusattrib}. The timestamp and CAN ID fields could be used to detect DoS and replay attacks, which exploits the vulnerability of ID priority as described previously in section \ref{sec:3}. The DLC and DATA fields could be used to detect falsified and injected frames. Then, the feature vector is transferred to the analyzer module. 

Both network and in-vehicle feature vectors are pre-processed before their transfer to the analyzer module. Pre-processing operations concern mainly removing missing values and scaling feature values using z-transformation. The pseudo-code algorithm \ref{alg:tcm} summarizes the different steps of the traffic collection module.

\begin{table}[!htbp]
\centering
\small
  \caption{CAN bus frames feature vector attributes}
  \label{tab:canbusattrib}
\begin{adjustbox}{width=0.4\textwidth}
\begin{tabular}{|l|l|}
\hline
\multicolumn{1}{|c|}{{\color[HTML]{000000} Features}} & \multicolumn{1}{c|}{{\color[HTML]{000000} Description}}    \\ \hline
{\color[HTML]{000000} Timestamp}                      & {\color[HTML]{000000} Recorded time (s)}                   \\ \hline
{\color[HTML]{000000} CAN ID}                         & {\color[HTML]{000000} Identifier of CAN message}           \\ \hline
{\color[HTML]{000000} DLC}                            & {\color[HTML]{000000} Number of data bytes, from 0 to 8}   \\ \hline
{\color[HTML]{000000} DATA{[}0{]}}                    & {\color[HTML]{000000} }                                    \\ \cline{1-1}
{\color[HTML]{000000} DATA{[}1{]}}                    & {\color[HTML]{000000} }                                    \\ \cline{1-1}
{\color[HTML]{000000} DATA{[}2{]}}                    & {\color[HTML]{000000} }                                    \\ \cline{1-1}
{\color[HTML]{000000} DATA{[}3{]}}                    & {\color[HTML]{000000} }                                    \\ \cline{1-1}
{\color[HTML]{000000} DATA{[}4{]}}                    & {\color[HTML]{000000} }                                    \\ \cline{1-1}
{\color[HTML]{000000} DATA{[}5{]}}                    & {\color[HTML]{000000} }                                    \\ \cline{1-1}
{\color[HTML]{000000} DATA{[}6{]}}                    & {\color[HTML]{000000} }                                    \\ \cline{1-1}
{\color[HTML]{000000} DATA{[}7{]}}                    & \multirow{-8}{*}{{\color[HTML]{000000} Data value (byte)}} \\ \hline
\end{tabular}
\end{adjustbox}
\end{table}

%%%%%%%%%%%%%%%%%%%%%%%%%%%%%%%%%%%%%%%%%%%%%%%%%%%%%%%%%%%%%%%%%
\begin{algorithm}[h!]
\small
%\color{blue}
    \caption{Traffic Collection Module}
    \label{alg:tcm}
     \textbf{BEGIN}\\
     \textbf{Input :} \textit{ $CBF$ (CAN bus frame), $NF$ (Network flow)}
    
    \textbf{Variables :}: \textit{ $NFw$ (Network flow feature vector), $CBFw$ (CAN bus frame feature vector)}

    \If{($NF$ is captured)}{
    	
    	 Extract Features from ($NF$); \\
    	 Generate ($NFw$);\\
    	 Preprocess ($NFw$);\\
    	 Send ($NFw$) to Analyzer Module;
    		}
  
   \If{($CBF$ is captured)}{
    	
    	 Extract Features from $CBF$; \\
    	 Generate ($CBFw$);\\
    	 Preprocess ($CBFw$);\\
    	 Send ($CBFw$) to Analyzer Module;

    		} 
  	 \textbf{END}
\end{algorithm}
%%%%%%%%%%%%%%%%%%%%%%%%%%%%%%%%%%%%%%%%%%%%%%%%%%%%%%%%%%%%%%%%%

\subsection{Analyzer module} \label{sec:analyzer} 
It is the most important module of the framework. Since this module handles two types of independent traffic, it uses two analyzers. The first analyzer is responsible for analyzing network traffic, the second one to analyze CAN bus frames. The network analyzer is a classifier trained using supervised machine learning algorithms on legitimate and malicious network traffic. The legitimate network traffic is generated by running specialized vehicular application related to safety \cite{gelbal2019cooperative}, convenience \cite{ta2020secure} and commercial \cite{liu2011remote} applications. The malicious network traffic is related to typical bot malware's activities such as DOS and information theft attacks. The in-Vehicle analyzer is a classifier trained using supervised machine learning algorithms on legitimate and malicious (DoS, Fuzzy, RPM, and Gear) CAN frames. Both classifiers are generated in a central server then integrated into the framework. Within the vehicle, the two classifiers are used for continuous monitoring of network traffic and CAN bus traffic. 

The network analyzer uses the calculated network features to classify the received network flow in one of three classes: normal, DOS, or information theft. The in-Vehicle analyzer classifies CAN bus frames based on the calculated features into four classes: normal, DOS, frames injection or Replay attack. If a malicious network flow or CAN bus frame is detected, then the analyzer module sends an alert to the manager module to take immediate action. Otherwise, it will be ignored.

To train both analyzers we use the following supervised machine learning algorithms. For each analyzer, we choose the algorithm that gives the best performances. The following algorithms are selected for their known efficiency and classification performances:
%[start=1,label={\bfseries 5.3.\arabic*}]
\begin{enumerate}  
\item \textbf{Naive Bayes Algorithm: } It is a probabilistic classifier that makes classifications using the maximum a posteriori decision rule in a Bayesian setting. It operates on a strong independence assumption, which means that the probability of one attribute does not affect the probability of the other.
\begin{equation}
P(c\mid x)=\frac{{P(x \mid c)}*P(c)}{P(x)}
\label{naiv}   
\end{equation}
\begin{equation}
P(c\mid x)=P(x1\mid c)*P(x2\mid c)*...*P(xn\mid c)*P(c)
\label{naiv2}
\end{equation}

$P(c|x)$ is the posteriori probability of class (target) given predictor (attribute). 
$P(c)$ is the prior probability of class.
$P(x|c)$ is the likelihood, which is the probability of predictor given class.
$P(x)$ is the prior probability of predictor.

\item \textbf{Support Vector Machine(SVM):} Each data item is plotted as a point in n-dimensional space (where n is the number of features) with the value of each feature being the value of a particular coordinate. Then, the classification technique is performed to differentiate the classes and define which one the data points belong to.

\item \textbf{K-Nearest Neighbour: } the k-nearest neighbours algorithm (k-NN) is a non-parametric method used for classification and regression \cite{tan2005neighbor}. It works based on minimum distance from the query instance to the training samples to determine the K-nearest neighbours. After gathering the K nearest neighbours, it takes a simple majority of these K-nearest neighbours to be the prediction of the query instance.
Formally:
\begin{equation}
score (D, C_{i})= \sum_{D_{j} \in KNN_{d}}^{} Sim(D, D_{j})\wp (D_{j}, C_{i})
\label{knn}
\end{equation}

Above, \textit{KNN(d)} indicates the set of K-nearest neighbours of the query instance
\begin{equation}
D* \wp (Dj, Ci)
\label{knn2}
\end{equation}
with respect to class \textit{Ci}, that is:
\begin{equation}
\wp (Dj, Ci)=\begin{cases}1, Dj \in Ci \\- 0, Dj \notin Ci   \end{cases}
\label{knn3}
\end{equation}
\textit{Sim(D, Dj)} : represent the similarity score of each nearest neighbour document to the test document. Moreover, it is used as the weight of the classes of the neighbour document.

For test document \textit{d}, it should be assigned the class that has the highest resulting weighted sum.

\item \textbf{Decision Trees: } to build a decision tree, in this paper we used the ID3 (Iterative Dichotomiser 3) algorithm that uses entropy and information gain as metrics. Entropy characterizes the impurity of an arbitrary collection of examples, it can be defined formally as follow:
\begin{equation}
H(s)=\sum_{c\in C}-P(c)*log_{2} P(c)
\label{decisiontree}
\end{equation}
where \textit{S} is the current data set for which entropy is being calculated, \textit{C} is the set of classes in \textit{S}, and \textit{P(c)} represents the proportion of elements in class c to the number of elements in set S.

\item \textbf{Random Forest: }like its name implies, consists of a large number of individual decision trees that operate as an ensemble. Each individual tree in the random forest spits out a class prediction and the class with the most votes becomes our model’s prediction.

\item \textbf{Neural Networks: }is one of the most known machines learning algorithms. It works based on several layers to analyze the data. Each layer tries to detect patterns on the input data. When one of the patterns is detected, the next hidden layer is activated and so on \cite{pelk_2017}. In this paper, we use Multilayer perceptron (MLP), a supplement of feed-forward artificial neural network. It consists of three types of layers: the input layer, output layer and hidden layer. The input layer receives the input signal to be processed. The required task such as prediction and classification is performed by the output layer. An arbitrary number of hidden layers that are placed in between the input and output layer is the true computational engine of the MLP \cite{abirami2020energy}.
\end{enumerate}

The pseudo codes algorithm \ref{alg:am} summarizes the different steps of the analyzer module.

%%%%%%%%%%%%%%%%%%%%%%%%%%%%%%%%%%%%%%%%%%%%%%%%%%%%%%%%%%%%%%%%
\begin{algorithm}[h!]
\small
%\color{blue}
    \caption{Analyzer Module}
    \label{alg:am}
    
     \textbf{BEGIN}\\
     \textbf{Input :} \textit{ $CBFw$ (CAN bus frame feature vector), $NFw$ (Network flow feature vector)}
     
     \textbf{Output :} \textit{ $CBFw\_Decision$ (CAN bus frame feature vector Decision), $NFw\_Decision$ (Network flow feature vector Decision)}

    \If{($NFw$ is Received)}{
    	   $NFw\_Decision$ <- Network\_Analyzer\_SubModule ($NFw$);\\
    	   Send ($NFw$, $NFw\_Decision$) to Manager Module;
    		}
    \If{($CBFw$ is Received)}{
    	   $CBFw\_Decision$ <- In-Vehicle\_Analyzer\_SubModule ($CBFw$);\\
    	   Send ($CBFw$, $CBFw\_Decision$) to Manager Module;
    		}
 
  	 \textbf{END}\\

%%%%%%%%%%%%% Network Analyzer SubModule %%%%%%%%%%%%%%%

\SetKwFunction{FMain}{Network\_Analyzer\_SubModule}
    \SetKwProg{Fn}{Function}{:}{}
    \Fn{\FMain{$N$}}{
    
    \textbf{Local Variable :} \textit{ $N\_Prediction$ (Network Traffic Prediction)}\\
    
    \textbf{$N\_Prediction$} = Machine Learning Algorithm($N$);\\
    
    \textbf{return} $N\_Prediction$;
}
\textbf{End Function}
%%%%%%%%%%%%%%%%%%%%%%%%%%%%%%%%%%%%%%%%%%%%%%%%%%%%%

%%%%%%%%%%% IN-Vehicle Analyzer SubModule %%%%%%%%%%%

\SetKwFunction{FMain}{In-Vehicle\_Analyzer\_SubModule}
    \SetKwProg{Fn}{Function}{:}{}
    \Fn{\FMain{$F$}}{
    
    \textbf{Local Variable :} \textit{ $F\_Prediction$ (Frame Prediction)}\\
    
    \textbf{$F\_Prediction$} = Machine Learning Algorithm($F$);\\
    
    \textbf{return} $F\_Prediction$;
}
\textbf{End Function}
\end{algorithm}

\subsection{Manager module}
The manager module handle alerts and triggers adequate response measures according to the detected attacks. Whenever the analyzer module detects a botnet activity, it sends an alert to the manager module. The latter logs the traces of the corresponding event and notify the driver. If a DoS attack is detected the manager module terminates the network session with the victim vehicle. In the case of theft of information, the manager saves logs and notifies the driver. If the driver does not approve the transfer, the connection to the destination address will be blocked. Otherwise, the flow will be ignored. At the in-vehicle level, when the analyzer detects a CAN frame as belonging to a botnet activity, it sends an alert to the manager, which will notify the driver. To avoid interrupting wrongly vehicle's services, whatever the type of detected attacks, the manager module asks for the driver's approval before undertaking any response measures. When the driver is notified depends on the attack. The most serious cases (disabling brake attacks) require direct notification. However, in other scenarios that do not have an impact on the driver's life, the notification is not done while driving, but in the moments when the vehicle is completely stopped, in order to avoid unnecessary shocks. The pseudo-codes algorithm \ref{alg:sh} summarizes the different responses measures.

Terminating malicious processes running on the vehicle does not mean that it will not be executed again. To get rid of this inappropriate malware that was detected, the system of the vehicle must be reset. The resetting process removes the applications and files installed on the system, which might be the carriers of the malware. The manager module has to ask the permission of the driver before the reset operation.

%%%%%%%%%%%%%%%%%%%%%%%%%%%%%%%%%%%%%%%%%%%%%%%%%%%%%%%%%%%%%%%%%
\begin{algorithm}[h!]
\small
%\color{blue}
    \caption{Manager module response measures}
    \label{alg:sh}
     \textbf{BEGIN}
     
    \textbf{Input}: \textit{ $CBFw$ (CAN bus frame feature vector),$NFw$ (Network flow feature vector), $CBFw\_Decision$ (CAN bus frame feature vector Decision), $NFw\_Decision$ (Network flow feature vector Decision)}
    
    \If{($CBFw\_Decision$ classified as malicious)}{
    	
    	 Send an alert to the driver about $CBFw$ \\
    	 Ask driver's approval for system reset
    	 
    		}

   \eIf{($NFw\_Decision$ classified as DoS attack)}{
    	
    	 Save detailed logs of $NFw$ \\
    	 Send alert to the driver\\
    	 Terminate network session \\
    	 Ask driver's approval for system reset
    	 
    		} { \If{($NFw\_Decision$ classified as theft of information)}{
    	
    	 Save detailed logs of $NFw$ \\
    	 Send an alert to the driver\\ 
    	 {\eIf{(information theft confirmed)}{
    	
    	 Terminate network session \\
    	 Ask driver's approval for system reset \\
    		} {Ignore $NFw$ \;  }}
    		} 
    		
    		         { Ignore $NFw$\;}   
    		        
    		        }

  	 \textbf{END}
\end{algorithm}

Finally, to keep the trained model up to date (against new zero-day malware attacks), a new training session is started automatically whenever the accuracy of the model starts degrading. The management module starts sending the incoming traffic collected by the collection module to a remote server in the cloud, which is used to perform exhaustive computation operations. The traffic data sent to the cloud server contains all fields defined by the network protocols, but not whether the traffic is malicious or not (unlabelled data). Therefore, the first step is to label the data. To ensure an automatic labelling method and avoid the manual intervention, we use an ensemble method, which classifies new samples by taking the majority vote of several classifiers (decision tree, random forest, and neural networks algorithms). Ensemble methods generally provide more accurate results than a single classifier and can reduce the rate of labelling errors. The labelling operation is done at the server level and not at the vehicle level to avoid computationally intensive operations and the memory requirements caused by using several classification algorithms together in the voting process. In a second step, we combine the newly collected (and labelled) data with the old training data to update the database used to train our classifier. In a third step, the analysis module uses the new database for retraining, so that it can be able to detect the new malware traffic and provide an appropriate result. The advantage of such an update technique is that it is fully automated.

%To do this automatically, we use a voting method, which classifies new samples by taking the majority vote of several classifiers (example: Decision Tree, Random Forest, and Neural Networks algorithms).

\section{Experimentation} \label{sec:5}
In this section, we provide the evaluation results of the proposed framework. First, we describe in details the datasets used in this research, and the pre-processing operations we have carried out. Since the proposed framework monitors the in-vehicle and network communication, we have used two datasets to evaluate its performance, the first one contains vehicular network traffic, the second one in-vehicle traffic \cite{seo2018gids}. A detailed description of the two datasets is provided below.

\begin{table*}[!h]
  \caption{Vehicular Networks applications}
  \label{tab:VANETNetApp}
\begin{adjustbox}{width=1.0\textwidth, center}
\begin{tabular}{|c|c|l|}
\hline
Type                                                                            & Abbreviation & \multicolumn{1}{c|}{Description}                                                                                                                                                                                             \\ \hline
\multirow{8}{*}{\begin{tabular}[c]{@{}c@{}}safety\\ oriented\end{tabular}}      & BSM          & \begin{tabular}[c]{@{}l@{}}Basic Safety Message: a packet transmitted approximately 10 times per second. It contains \\ vehicle's state information, like speed, GPS coordination, … etc.\end{tabular}                       \\ \cline{2-3} 
                                                                                & CCW          & \begin{tabular}[c]{@{}l@{}}Cooperative Collision Warning: an application that aims to help the driver to avoid collisions\\ by giving him the kinematics status messages collected from the vehicles around.\end{tabular}        \\ \cline{2-3} 
                                                                                & CVW          & \begin{tabular}[c]{@{}l@{}}Cooperative Violation Warning: roadside units send to drivers the necessary information \\ when approaching to a signal phase or a red light.\end{tabular}                                        \\ \cline{2-3} 
                                                                                & EEBL         & \begin{tabular}[c]{@{}l@{}}Emergency Electronic Brake Light: If a vehicle braking hard, it broadcasts a message to inform\\ the other vehicles.\end{tabular}                                                                         \\ \cline{2-3} 
                                                                                & PCN          & \begin{tabular}[c]{@{}l@{}}V2V Post Crash Notification: If a vehicle is involved in a crash, it broadcasts a message to inform\\ the other vehicles.\end{tabular}                                                                    \\ \cline{2-3} 
                                                                                & RFN          & \begin{tabular}[c]{@{}l@{}}Road Feature Notification: if a vehicle detects an advisory road feature (acute turn), it broadcasts\\ a message to inform the other vehicles.\end{tabular}                                               \\ \cline{2-3} 
                                                                                & RHCN         & \begin{tabular}[c]{@{}l@{}}Road Hazard Condition Notification: if a vehicle detects a road hazard (rocks, animals, ice), it \\ broadcasts a message to inform the other vehicles.\end{tabular}                                       \\ \cline{2-3} 
                                                                                & SVA          & \begin{tabular}[c]{@{}l@{}}Stopped or Slow Vehicle Advisor: If a vehicle reduces its speed or stops, it broadcasts a message\\ to inform the other vehicles.\end{tabular}                                                            \\ \hline
\multirow{5}{*}{\begin{tabular}[c]{@{}c@{}}convenience\\ oriented\end{tabular}} & CRN          & \begin{tabular}[c]{@{}l@{}}Congested Road Notification: If a vehicle detects a collision, it sends a message to alert the\\ other vehicles so that they can take another lane.\end{tabular}                                      \\ \cline{2-3} 
                                                                                & PAN          & \begin{tabular}[c]{@{}l@{}}Parking Availability Notification: this application is used to request the road side unit to \\ provide the closest parking and if they contain free parking spaces.\end{tabular}                 \\ \cline{2-3} 
                                                                                & PSL          & \begin{tabular}[c]{@{}l@{}}Parking Spot Locator: It is an exchange of information between a vehicle and the Parking \\ Space Locator roadside unit. It is used to provide free parking places in a parking lot.\end{tabular} \\ \cline{2-3} 
                                                                                & TOLL         & Free Flow Tolling: it permits to apply an e-payment while passing through a highway toll gates.                                                                                                                              \\ \cline{2-3} 
                                                                                & TP           & \begin{tabular}[c]{@{}l@{}}Traffic Probe: The kinematics status messages are collected and transmitted through road side\\ units to the traffic management center.\end{tabular}                                              \\ \hline
\multirow{4}{*}{\begin{tabular}[c]{@{}c@{}}commercial\\ oriented\end{tabular}}  & CMDD         & \begin{tabular}[c]{@{}l@{}}Content, Map or Database Download: a vehicle connect to a wireless hot-spot to download\\ content (maps, multimedia, or web pages).\end{tabular}                                                  \\ \cline{2-3} 
                                                                                & RTVR         & \begin{tabular}[c]{@{}l@{}}Real-time Video Relay: a vehicle may initiate transfer of real-time video that may be \\ useful to other drivers in the area.\end{tabular}                                                    \\ \cline{2-3} 
                                                                                & RVP/D        & \begin{tabular}[c]{@{}l@{}}Remote Vehicle Personalization/Diagnostics: the drivers can connect to a hot-spot to download\\  the latest personalized vehicle settings.\end{tabular}                                           \\ \cline{2-3} 
                                                                                & SA           & \begin{tabular}[c]{@{}l@{}}Service Announcement: Fast food or restaurant use an infrastructure that send periodically \\ Hello messages to announce the vehicles of its presence.\end{tabular}                               \\ \hline
\end{tabular}
\end{adjustbox}
\end{table*}

\subsection{Network traffic dataset}

\begin{figure}[h!]
    \centering
    \includegraphics[width=0.8\columnwidth, height=0.9\columnwidth]{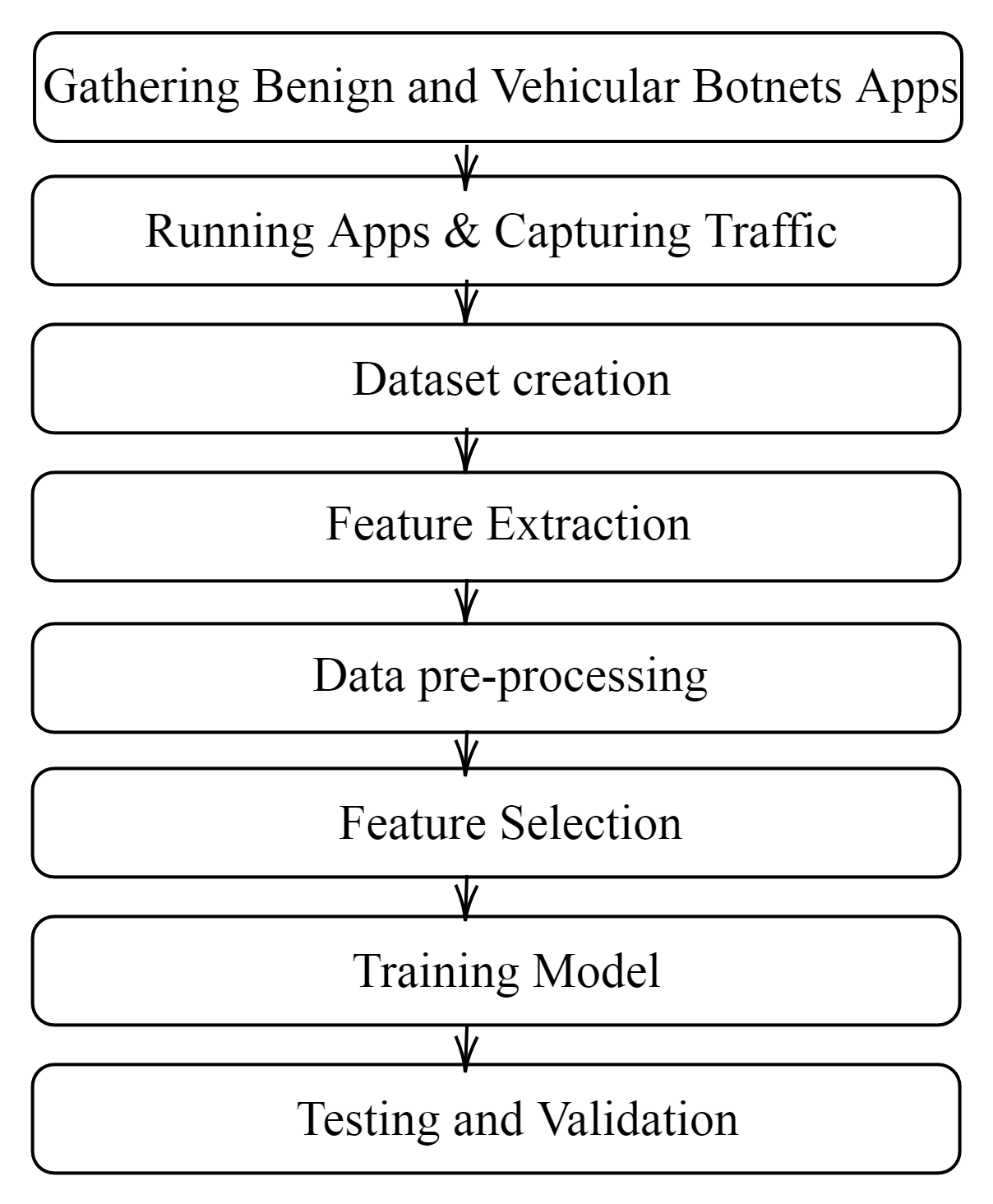}
    \caption{Dataset generation flow chart}
    \label{fig:my_label}
\end{figure}

\subsubsection{Generating network traffic data}\label{sec:5.1.1}

To the best of our knowledge, no real vehicular network traffic dataset including botnet traffic is publicly available. Thus, we simulate vehicular botnet activity discussed in section \ref{sec:3}. To generate realistic benign vehicular network traffic, we have implemented 17 applications including safety, convenience and commercial applications, the list of applications (inspired from \cite{krishnan2010commercial}) with their brief description is provided in table \ref{tab:VANETNetApp}, and the simulation parameters and scenarios are described further in this section. To ensure that the generated traffic is representative of real benign traffic and covers diverse types of application, we have considered the following factors for choosing applications: 1) physical-layer channel (CCH \& SCH); 2) transfer protocols (IP \& WSMP); 3) message TTL (single-hop \& multi-hop); 4) routing protocol (geocast, broadcast \& unicast); 5) trigger Condition (On-demand \& event-triggered); 6) and communication technology (V2V \& V2I). Table \ref{tab:VANETAppNetworkAtt} provides a categorization of the benign applications based on the aforementioned factors.

To generate malicious network traffic related to a bot malware activity, we have simulated the following bot activity scenarios: 1) WSMP flood; 2) geo WSMP Flood; 3) GPS tracking; 4) and information theft, through eavesdropping on drivers and passenger's conversations.

To simulate the aforementioned benign vehicular network applications and the bot malware activity, we implemented different scenarios based on nodes ID. Four scenarios to simulate the bot activity, and six scenarios to simulate the benign vehicular network traffic. In every scenario, the simulation runs for 500 seconds with 40 nodes moving according to the random waypoint model with a speed of 20 m/s and no pause time within the Manhattan map (downloaded it from OpenStreetMap \cite{openstreetmap}). The WiFi is 802.11p, the transmit power is set to 20 dBm, and the default propagation model is Two-Ray Ground.

The first and second scenarios correspond to zero-day DDoS attacks applicable only in vehicular networks environment (highlighted earlier in section \ref{sec:3}). In these scenarios, the victim nodes are node 0, node 6, and node 10; while the other nodes are the attackers. As regards the GPS tracking and theft of information scenarios, node 20 represent the botmaster, and nodes 7, 8,  and 9 are the victims. In the GPS tracking scenario, victim nodes send theirs GPS information to the botmaster in real-time, while in the information theft scenario, victim nodes send the recorded audio files periodically. For the benign traffic scenarios, we have simulated in each scenario the applications that have common features according to their characteristics, and in which the 40 nodes participate.

We used Network Simulator version 3 (NS3) \cite{nsnam} and the Simulation of Urban MObility (SUMO) package \cite{sumo}. SUMO is responsible for simulating realistic vehicular traffic, while NS3 is used to simulate the communication capabilities of the vehicles with IEEE 802.11p integration. Table \ref{tab:simpar} summarizes the simulation parameters, and table \ref{tab:samples} presents the samples distribution of the network traffic dataset.

\def\checkmark{\tikz\fill[scale=0.4](0,.35) -- (.25,0) -- (1,.7) -- (.25,.15) -- cycle;}

\begin{table*}[!htbp]
  \caption{Vehicular networks applications classified based on network attributes}
  \label{tab:VANETAppNetworkAtt}
\begin{adjustbox}{width=1.0\textwidth, center}
\begin{tabular}{|c|c|c|c|c|c|c|c|c|c|c|c|c|c|c|c|c|}
\hline
\multicolumn{2}{|c|}{\multirow{2}{*}{Application}}                                      & \multicolumn{2}{c|}{Channel} & \multicolumn{2}{c|}{Protocol} & \multicolumn{2}{c|}{Message TTL} & \multicolumn{3}{c|}{Routing Protocol} & \multicolumn{3}{c|}{\begin{tabular}[c]{@{}c@{}}Trigger\\   Condition\end{tabular}}    & \multicolumn{3}{c|}{Participants} \\ \cline{3-17} 
\multicolumn{2}{|c|}{}                                                                  & CCH           & SSH          & WSMP           & IP           & Multi-hop      & Single-Hop      & Geocast    & Broadcast    & Unicast   & Beaconing & \begin{tabular}[c]{@{}c@{}}Event-\\     triggred\end{tabular} & On-demand & V2V      & V2I     & Internet     \\ \hline
\multirow{8}{*}{\begin{tabular}[c]{@{}c@{}}safety\\ oriented\end{tabular}}      & BSM   & \checkmark             & \xmark            &\checkmark             &\xmark           &\xmark             &\checkmark              &\xmark         &\checkmark           &\xmark        &\checkmark        &\xmark                                                            &\xmark        &\checkmark       &\xmark      &\xmark           \\ \cline{2-17} 
                                                                                & CCW   &\checkmark            &\xmark           &\checkmark             &\xmark           &\xmark             &\checkmark              &\xmark         &\checkmark           &\xmark        &\checkmark        &\xmark                                                            &\xmark        &\checkmark       &\xmark      &\xmark           \\ \cline{2-17} 
                                                                                & CVW   &\checkmark            &\xmark           &\checkmark             &\xmark           &\xmark             &\checkmark              &\xmark         &\checkmark           &\xmark        &\checkmark        &\xmark                                                            &\xmark        &\xmark       &\checkmark      &\xmark           \\ \cline{2-17} 
                                                                                & EEBL  &\checkmark            &\xmark           &\checkmark             &\xmark           &\checkmark             &\xmark              &\checkmark         &\xmark           &\xmark        &\xmark        &\checkmark                                                            &\xmark        &\checkmark       &\xmark      &\xmark           \\ \cline{2-17} 
                                                                                & PCN   &\checkmark            &\xmark           &\checkmark             &\xmark           &\checkmark             &\xmark              &\checkmark         &\xmark           &\xmark        &\xmark        &\checkmark                                                            &\xmark        &\checkmark       &\xmark      &\xmark           \\ \cline{2-17} 
                                                                                & RFN   &\checkmark            &\xmark           &\checkmark             &\xmark           &\checkmark             &\xmark              &\checkmark         &\xmark           &\xmark        &\xmark        &\checkmark                                                            &\xmark        &\checkmark       &\xmark      &\xmark           \\ \cline{2-17} 
                                                                                & RHCN  &\checkmark            &\xmark           &\checkmark             &\xmark           &\checkmark             &\xmark              &\checkmark         &\xmark           &\xmark        &\xmark        &\checkmark                                                            &\xmark        &\checkmark       &\xmark      &\xmark           \\ \cline{2-17} 
                                                                                & SVA   &\checkmark            &\xmark           &\checkmark             &\xmark           &\checkmark             &\xmark              &\checkmark         &\xmark           &\xmark        &\xmark        &\checkmark                                                            &\xmark        &\checkmark       &\xmark      &\xmark           \\ \hline
\multirow{5}{*}{\begin{tabular}[c]{@{}c@{}}convenience\\ oriented\end{tabular}} & CRN   &\xmark            &\checkmark           &\checkmark             &\xmark           &\checkmark             &\xmark              &\checkmark         &\xmark           &\xmark        &\xmark        &\checkmark                                                            &\xmark        &\checkmark       &\xmark      &\xmark           \\ \cline{2-17} 
                                                                                & PAN   &\xmark            &\checkmark           &\checkmark             &\xmark           &\checkmark             &\xmark              &\xmark         &\xmark           &\checkmark        &\xmark        &\xmark                                                            &\checkmark        &\xmark       &\checkmark      &\xmark           \\ \cline{2-17} 
                                                                                & PSL   &\xmark            &\checkmark           &\checkmark             &\xmark           &\xmark             &\checkmark              &\xmark         &\xmark           &\checkmark        &\xmark        &\xmark                                                            &\checkmark        &\xmark       &\checkmark      &\xmark           \\ \cline{2-17} 
                                                                                & TOLL  &\xmark            &\checkmark           &\checkmark             &\xmark           &\xmark             &\checkmark              &\xmark         &\xmark           &\checkmark        &\xmark        &\checkmark                                                            &\xmark        &\xmark       &\checkmark      &\xmark           \\ \cline{2-17} 
                                                                                & TP    &\xmark            &\checkmark           &\checkmark             &\xmark           &\checkmark             &\xmark              &\xmark         &\xmark           &\checkmark        &\xmark        &\checkmark                                                            &\xmark        &\xmark       &\checkmark      &\xmark           \\ \hline
\multirow{4}{*}{\begin{tabular}[c]{@{}c@{}}commercial\\ oriented\end{tabular}}  & CMDD  &\xmark            &\checkmark           &\xmark             &\checkmark           &\xmark             &\checkmark              &\xmark         &\xmark           &\checkmark        &\xmark        &\xmark                                                            &\checkmark        &\xmark       &\xmark      &\checkmark           \\ \cline{2-17} 
                                                                                & RTVR  &\xmark            &\checkmark           &\xmark             &\checkmark           &\checkmark             &\xmark              &\xmark         &\xmark           &\checkmark        &\xmark        &\xmark                                                            &\checkmark        &\xmark       &\xmark      &\checkmark           \\ \cline{2-17} 
                                                                                & RVP/D &\xmark            &\checkmark           &\xmark             &\checkmark           &\xmark             &\checkmark              &\xmark         &\xmark           &\checkmark        &\xmark        &\xmark                                                            &\checkmark        &\xmark       &\checkmark      &\xmark           \\ \cline{2-17} 
                                                                                & SA    &\xmark            &\checkmark           &\xmark             &\checkmark           &\xmark             &\checkmark              &\checkmark         &\xmark           &\xmark        &\xmark        &\xmark                                                            &\checkmark        &\xmark       &\xmark      &\checkmark           \\ \hline
\end{tabular}
\end{adjustbox}
\end{table*}

\subsubsection{Features extraction}
After collecting the network traffic generated during simulation (as PCAP files), we have extracted a set of 79 network features (see table \ref{tab:features_network}). To extract features we have used CICFlowMeter \cite{website3}, a network traffic flow generator distributed by the Canadian Institut for Cyber Security (CIC). It generates bidirectional flows, where the first packet determines the forward (source to destination) and backward (destination to source) directions. Note that TCP flows are usually terminated upon connection teardown (by FIN packet) while UDP flows are terminated by a flow timeout. The flow timeout value can be assigned arbitrarily by the individual scheme e.g., 600 seconds for both TCP and UDP.

The list of network flow-based features extracted for each network flow is composed of three categories of features: time, bytes, and packets based features. We believe that time-based features (Flow IAT, Fwd IAT and Idle Time) are useful to detect DoS attacks because the time interval between successive packets is too short. Also, time-based features allow detection of periodic events such as periodic transfer of collected information in the case of theft of information. The bytes/packets based features allow the detection of large and abnormal traffic increases, which are symptomatic of DoS attacks.

\begin{table}[!htbp]
  \caption{Simulation parameters}
  \label{tab:simpar}
\centering
\begin{adjustbox}{width=0.45\textwidth}
\begin{tabular}{|l|l|}
\hline
Network Simulator  & NS3                                    \\ \hline
Traffic Generator  & SUMO                                   \\ \hline
Simulation Area    & Manhattan  Map                         \\ \hline
Simulation Time    & 500 seconds                            \\ \hline
Number of Nodes    & 40                                     \\ \hline
Max Speed          & 20 m/s                                 \\ \hline
MAC/PHY Standard   & IEEE802.11p                            \\ \hline
Traffic Type       & WSMP, IP                               \\ \hline
Bandwidth Channel  & CCH, SCH                               \\ \hline
Propagation Model  & Two-ray ground-reflection model        \\ \hline
Transmission Power & 20 dBm                                 \\ \hline
Packets Size       & Depend on the application and protocol \\ \hline
Packets Data Rate  & Depend on the application              \\ \hline
\end{tabular}
\end{adjustbox}
\end{table}

\begin{table*}[!htbp]
  \caption{Statistics of normal and attack samples of network traffic dataset}
  \label{tab:samples}
\begin{adjustbox}{width=0.7\textwidth, center}
\begin{tabular}{cccccccc}
\hline
\multicolumn{2}{c}{\multirow{2}{*}{Samples}} & \multicolumn{3}{c}{Nb} & \multicolumn{3}{c}{\%}      \\ \cline{3-8} 
\multicolumn{2}{c}{}                         & Train  & Test  & Total & Train   & Test    & Total   \\ \hline
\multicolumn{2}{c}{Benign}                   & 909    & 606   & 1515  & 31.14\% & 20.76\% & 51.90\% \\ \hline
\multirow{4}{*}{Malicious} & GPS Tracking    & 295    & 197   & 492   & 10.11\% & 6.74\%  & 16.85\% \\
                           & Phishing Attack & 191    & 128   & 319   & 6.55\%  & 4.37\%  & 10.92\% \\
                           & WSMP-Flood      & 182    & 121   & 303   & 6.23\%  & 4.15\%  & 10.38\% \\
                           & Geo-WSMP-Flood  & 174    & 116   & 290   & 5.96\%  & 3.97\%  & 9.93\%  \\ \hline
\multicolumn{2}{c}{Total}                    & 1751   & 1168  & 2919  & 60\%    & 40\%    & 100\%   \\ \hline
\end{tabular}
\end{adjustbox}
\end{table*}

\subsubsection{Data pre-processing and features selection}
For the data pre-processing step, we did the cleaning and the normalization. To check missing values and deal with them, we used one of the python programming language functions named Dropna(). Dropna removes a row or a column from a data frame, which has a NaN or no values in it. Moreover, to deal with the huge differences between magnitude, units, and range in the generated dataset, we used feature scaling. The feature scaling aims to put all the values in the data set between 0 and 1, to make the features more consistent with each other and to make the training step-less sensitive to this problem.

We have used in this research two types of feature selection algorithms. The first one is forward selection \cite{websiteForwardSelection}, which belongs to the wrappers class. Forward selection is an iterative algorithm that starts with an empty set of feature. In each iteration, it adds the best feature that improves the model until the addition of a new feature does not improve the performance of the model. The Forward feature selection algorithm has reduced the number of features from 37 to 18 features as shown in table \ref{tab:featuresSelectionAlgorithms}. The second feature selection algorithm is the Linear Support Vector Classifier LinearSVC \cite{websiteLinearSVC}, which belongs to the embedded features selection algorithms. LinearSVC is an algorithm that gives to each feature a coef\_ or feature\_importances\_ attribute. All the features are considered unimportant at the beginning and it gives them values under a threshold parameter. After that, it uses built-in heuristics for finding the threshold of every feature using a string argument. LinearSVC has reduced the number of features from 37 to 22 as shown in table \ref{tab:featuresSelectionAlgorithms}. The best results were achieved when using the subset of features giving by the Forward selection algorithm (as shown in the results section).

%%%%%%%%%%%%%%%%%%%%%%%%%%%%%%%%%%%%%%%%%%%%%%%%%%%%%%%%%%%%%%%%%%%%%%%%%%%%%%%%%%%%%%%%%%%%%%%%%%%%%%%%%%%%%%%%%%%%%%%%%%

\begin{table*}[!htbp]
  \caption{List of selected network features}
  \label{tab:featuresSelectionAlgorithms}
\centering
\begin{adjustbox}{max width=0.7\textwidth}
\begin{tabular}{|c|c|}
\hline
Forward selection features set                                                                                                                                                                                                                                                                    & LinearSVC features set                                                                                                                                                                                                                                                                                                                                                             \\ \hline
\begin{tabular}[c]{@{}c@{}}Flow Duration, Tot Fwd Pkts, Flow Pkts/s, \\ Flow IAT Mean, Flow IAT Std, Flow IAT Max, \\ Flow IAT Min, Fwd IAT Tot, Fwd IAT Mean, \\ Fwd IAT Std, Fwd IAT Max, Fwd IAT Min, \\ Fwd Pkts/s, Bwd Pkts/s, Down/Up Ratio, \\ Idle Mean, Idle Std, Idle Min.\end{tabular} & \begin{tabular}[c]{@{}c@{}}Tot Fwd Pkts, Tot Bwd Pkts, TotLen Fwd Pkts,\\ Flow Byts/s, Flow Pkts/s, Flow IAT Mean,\\ Flow IAT Std, Flow IAT Max, Flow IAT Min,\\ Fwd IAT Mean, Fwd IAT Std, Fwd IAT Max, \\ Fwd IAT Min, Fwd Header Len, Fwd Pkts/s,\\ Bwd Pkts/s, Pkt Size Avg, Bwd Blk Rate Avg,\\ Init Fwd Win Byts, Init Bwd Win Byts,\\ Active Mean, Active Max.\end{tabular} \\ \hline
\end{tabular}
\end{adjustbox}
\end{table*}

\subsection{In-vehicule traffic dataset}
We have used in our experimentation the dataset built in \cite{seo2018gids}. The authors used Hyundai’s YF Sonata as a testing vehicle. They connected a Raspberry Pi3 with CAN bus through the OBD-II port (Figure \ref{fig:invehiclenetwork}), and connect the Raspberry to a laptop computer through WiFi. They generated an in-Vehicle dataset that contains normal CAN frames and other malicious (DoS frames, Fuzzy attack, RPM and Gear). The dataset is available online \cite{In-hacking}, labelled and in CSV format. We carried the same preprocessing operation of cleaning and normalization described in the previous section. Table \ref{tab:samples2} presents samples distribution of the in-vehicle traffic dataset.

\begin{table*}[!htbp]
  \caption{Statistics of normal and attack samples of in-vehicle traffic dataset}
  \label{tab:samples2}
\begin{adjustbox}{width=0.7\textwidth, center}
\begin{tabular}{cccccccc}
\hline
\multicolumn{2}{c}{\multirow{2}{*}{Samples}} & \multicolumn{3}{c}{Nb} & \multicolumn{3}{c}{\%}      \\ \cline{3-8} 
\multicolumn{2}{c}{}                         & Train  & Test  & Total & Train   & Test    & Total   \\ \hline
\multicolumn{2}{c}{Benign}                   & 1239   & 826   & 2065  & 26.93\% & 17.95\% & 44.88\% \\ \hline
\multirow{4}{*}{Malicious}   & DoS Attack    & 504    & 336   & 840   & 10.95\% & 7.3\%   & 18.25\% \\
                             & Fuzzy Attack  & 304    & 203   & 507   & 6.61\%  & 4.40\%  & 11.01\% \\
                             & Gear Attack   & 264    & 176   & 440   & 5.74\%  & 3.82\%  & 9.56\%  \\
                             & RPM Attack    & 449    & 300   & 749   & 9.76\%  & 6.51\%  & 16.27\% \\ \hline
\multicolumn{2}{c}{Total}                    & 2760   & 1841  & 4601  & 60\%    & 40\%    & 100\%   \\ \hline
\end{tabular}
\end{adjustbox}
\end{table*}

\subsection{Results and discussion} \label{sec:8}
To build the classification models for both analyzers (network and in-vehicle), we apply the supervised machine learning algorithms described in section \ref{sec:analyzer} with their default paramaters. For each analyzer, we choose the algorithm that gives the best performances. We train, validate and test the two classification models separately (network and in-vehicle analyzer). From each dataset, we take 60\% of the dataset to train and validate the classification model through 10-fold cross-validation, and the remainder 40\% for testing the model. Six common metrics, accuracy, precision, recall, F1\_score, false fositive rate, and false negative rate have been selected to evaluate the classification performances, the aforementioned metrics can be calculated as follows:

\begin{equation}
Accuracy =(TP+TN)/(TP+FP+FN+TN)
\label{decisiontree}
\end{equation}
\begin{equation}
Precision =TP/(TP+FP)
\label{decisiontree}
\end{equation}
\begin{equation}
Recall =TP/(TP+FN)
\label{decisiontree}
\end{equation}
\begin{equation}
\resizebox{.9 \columnwidth}{!} 
{
    F1\_score =(2*(precision*recall))/(precision+recall)
}
\label{decisiontree}
\end{equation}
\begin{equation}
FPR =FP/(TN+FP)
\label{decisiontree}
\end{equation}
\begin{equation}
FNR =FN/(TP+FN)
\label{decisiontree}
\end{equation}

where TP, FP, TN and FN denote true positive, false positive, true negative, and false negative, respectively.

\subsubsection{Malicious network traffic detection}
\begin{figure*}[h!]
\centering
\begin{adjustbox}{width=0.98\textwidth}
\pgfplotsset{width=0.93\textwidth,height=0.4\textheight,compat=1.16}
\begin{tikzpicture}   
\begin{axis}[
    legend columns=-1,
    legend entries={KNN, Neural Networks, Decision Tree, Random  Forest, SVM, Naive Bayes},
    legend to name=named,
    legend style={at={(0.5,-0.09)},
    anchor=north,legend columns=-1},
    symbolic x coords={Accuracy, F1-Score},
    major tick length=0cm,
    xtick=data,
    ymin=75.0,
     enlarge x limits=0.5,
    enlarge y limits={upper,value=0.2},
    nodes near coords,
    ybar,
    every node near coord/.append style={rotate=90, anchor=west},
    bar width = 15pt,
]

\addplot 
coordinates {(Accuracy,98.7) (F1-Score,86.1)};

\addplot 		
coordinates {(Accuracy,97.3) (F1-Score,86.92)};

\addplot 		
coordinates {(Accuracy,99.4) (F1-Score,90.19)};

\addplot 		
coordinates {(Accuracy,99.3) (F1-Score,87.08)};

\addplot 		
coordinates {(Accuracy,97.2) (F1-Score,83.74)};

\addplot 		
coordinates {(Accuracy,87.7) (F1-Score,80.54)};

\end{axis}
\end{tikzpicture}% NO EMPTY LINE HERE!!!! 
\begin{tikzpicture}  
\begin{axis}[
    legend style={at={(0.5,-0.09)},
    anchor=north,legend columns=-1},
    symbolic x coords={FPR,FNR},
    major tick length=0cm,
    xtick=data,
    ymin=0.0,
    enlarge x limits=0.5,
    enlarge y limits={upper,value=0.2},
    nodes near coords,
    ybar,
    every node near coord/.append style={rotate=90, anchor=west},
    bar width = 15pt,
]

\addplot 
coordinates {(FPR,1.2) (FNR,14.1)}; 

\addplot 		
coordinates {(FPR,2.67) (FNR,13.16)};

\addplot 
coordinates {(FPR,0.36) (FNR,10.29)}; 

\addplot 		
coordinates {(FPR,0.53) (FNR,12.84)};

\addplot 
coordinates {(FPR,2.96) (FNR,16.86)}; 

\addplot 		
coordinates {(FPR,3.61) (FNR,15.01)};

\end{axis}
\end{tikzpicture}
\end{adjustbox}

\begin{adjustbox}{max width=0.8\textwidth}
\ref{named}
\end{adjustbox}
\captionsetup{justification=centering}
\caption{Network traffic binary classification results: Left - Accuracy and F1-score. Right - FPR and FNR}
\label{fig:NetRes}
\end{figure*}
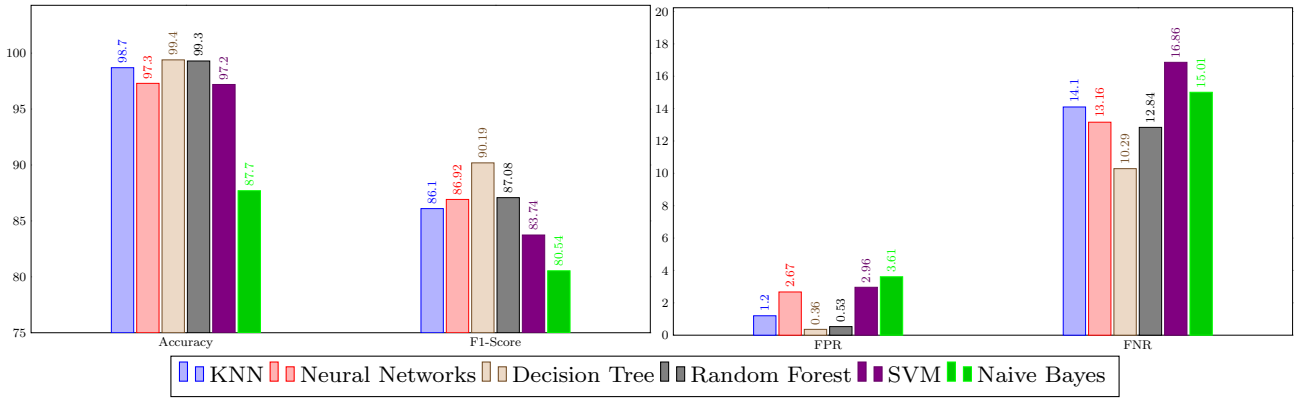

In this experiment, we evaluate the performances of AntibotV on classifying a network connection as legitimate or as malicious. As we can see from table \ref{tab:VANETNetRes}, all the classifiers show high accuracy ($>$ 85\%), the highest accuracy is achieved by decision tree (99.4\%) followed by random forest (99.3\%), while naive Bayes presents the lowest accuracy. However, accuracy is not enough to evaluate and select the best classification model. Therefore, other important metrics such as recall (detection rate), precision, false negative rate (FNR) and false positive rate (FPR) need to be taken into consideration. Figure \ref{fig:NetRes} compare the recall, F1-score, FNR, and FPR for benign and malicious traffic.

\begin{table*}[h!]
  \caption{Network traffic binary classification using AntibotV}
  \label{tab:VANETNetRes}
\small
\begin{adjustbox}{width=0.8\textwidth, center}
\begin{tabular}{cccccccc}
\hline
\multirow{2}{*}{Algorithm}       & \multirow{2}{*}{Classe} & \multirow{2}{*}{Precision} & \multirow{2}{*}{Recall} & \multirow{2}{*}{F1-score} & \multirow{2}{*}{FPR} & \multirow{2}{*}{FNR} & \multirow{2}{*}{Accuracy} \\
                                 &                         &                            &                         &                           &                      &                      &                           \\ \hline
\multirow{3}{*}{KNN}             & Benign traffic          & 98,5\%                     & 99,6\%                  & 99,0\%                    & 2,3\%                & 0,3\%                & \multirow{3}{*}{98,70\%}  \\
                                 & Malicious traffic       & 74,53\%                    & 72,15\%                 & 73,23\%                   & 0,05\%               & 27,83\%              &                           \\
                                 & Average Value           & 86,5\%                     & 85,9\%                  & 86,1\%                    & 1,2\%                & 14,1\%               &                           \\ \hline
\multirow{3}{*}{Neural Netowrks} & Benign traffic          & 97,20\%                    & 98,60\%                 & 97,90\%                   & 4,30\%               & 1,30\%               & \multirow{3}{*}{97,30\%}  \\
                                 & Malicious traffic       & 77,30\%                    & 74,94\%                 & 75,94\%                   & 1,04\%               & 25,02\%              &                           \\
                                 & Average Value           & 87,25\%                    & 86,77\%                 & 86,92\%                   & 2,67\%               & 13,16\%              &                           \\ \hline
\multirow{3}{*}{Decision Tree}   & Benign traffic          & 99,50\%                    & 99,60\%                 & 99,50\%                   & 0,70\%               & 0,30\%               & \multirow{3}{*}{99,40\%}  \\
                                 & Malicious traffic       & 82,88\%                    & 79,70\%                 & 80,88\%                   & 0,03\%               & 20,28\%              &                           \\
                                 & Average Value           & 91,19\%                    & 89,65\%                 & 90,19\%                   & 0,36\%               & 10,29\%              &                           \\ \hline
\multirow{3}{*}{Random Forest}   & Benign traffic          & 99,30\%                    & 99,50\%                 & 99,40\%                   & 1,00\%               & 0,40\%               & \multirow{3}{*}{99,30\%}  \\
                                 & Malicious traffic       & 74,80\%                    & 74,70\%                 & 74,75\%                   & 0,05\%               & 25,28\%              &                           \\
                                 & Average Value           & 87,05\%                    & 87,10\%                 & 87,08\%                   & 0,53\%               & 12,84\%              &                           \\ \hline
\multirow{3}{*}{SVM}             & Benign traffic          & 96,40\%                    & 99,40\%                 & 97,80\%                   & 5,80\%               & 0,50\%               & \multirow{3}{*}{97,20\%}  \\
                                 & Malicious traffic       & 74,28\%                    & 66,75\%                 & 69,68\%                   & 0,13\%               & 33,23\%              &                           \\
                                 & Average Value           & 85,34\%                    & 83,08\%                 & 83,74\%                   & 2,96\%               & 16,86\%              &                           \\ \hline
\multirow{3}{*}{Naive Bayes}     & Benign traffic          & 96,60\%                    & 83,60\%                 & 89,60\%                   & 4,50\%               & 16,30\%              & \multirow{3}{*}{87,70\%}  \\
                                 & Malicious traffic       & 65,95\%                    & 86,23\%                 & 71,48\%                   & 2,73\%               & 13,73\%              &                           \\
                                 & Average Value           & 81,28\%                    & 84,91\%                 & 80,54\%                   & 3,61\%               & 15,01\%              &                           \\ \hline
\end{tabular}
\end{adjustbox}
\end{table*}

%%%%%%%%%%%%%%%%%%%%%%%%%%%%%%%%%%%%%%%%%%%%%%%%%%%%%%%%%%%%%%%%

%%%%%%%%%%%%%%%%%%%%%%%%%%%%%%%%%%%%%%%%%%%%%%%%%%%%%%%%%%%%%%%%
\begin{table*}[h!]
\small
  \caption{Network traffic multiclass results: AntibotV based on Decision Tree classifier}
  \label{tab:valiceRandomRes}
\begin{adjustbox}{width=0.8\textwidth, center}
\begin{tabular}{cccccccc}
\hline
\multicolumn{2}{c}{\multirow{2}{*}{Classes}} & \multirow{2}{*}{Precision} & \multirow{2}{*}{Recall} & \multirow{2}{*}{F1-score} & \multirow{2}{*}{FPR} & \multirow{2}{*}{FNR} & \multirow{2}{*}{Accuracy} \\
\multicolumn{2}{c}{}                         &                            &                         &                           &                      &                      &                           \\ \hline
\multirow{2}{*}{Benign}    & WSMP Traffic    & 99,06\%                    & 97,70\%                 & 98,38\%                   & 0,16\%               & 2,29\%               & \multirow{7}{*}{99,40\%}  \\
                           & IP Traffic      & 99,57\%                    & 99,71\%                 & 99,64\%                   & 0,39\%               & 0,28\%               &                           \\ \cline{1-7}
\multirow{4}{*}{Malicious} & GPS Tracking    & 99,61\%                    & 100,00\%                & 99,80\%                   & 0,08\%               & 0,00\%               &                           \\
                           & Phishing Attack & 99,47\%                    & 100,00\%                & 99,73\%                   & 0,07\%               & 0,00\%               &                           \\
                           & WSMP Flood      & 97,43\%                    & 98,70\%                 & 98,06\%                   & 0,14\%               & 1,29\%               &                           \\
                           & GeoFlood        & 100,00\%                   & 91,66\%                 & 95,65\%                   & 0,00\%               & 8,30\%               &                           \\ \cline{1-7}
\multicolumn{2}{c}{Average Value}            & 99,19\%                    & 97,96\%                 & 98,54\%                   & 0,14\%               & 2,03\%               &                           \\ \hline
\end{tabular}
\end{adjustbox}
\end{table*}
%%%%%%%%%%%%%%%%%%%%%%%%%%%%%%%%%%%%%%%%%%%%%%%%%%%%%%%%%%%%%%%%
\begin{figure*}[h!]
\centering
\begin{adjustbox}{width=0.98\textwidth}
\pgfplotsset{width=0.93\textwidth,height=0.4\textheight,compat=1.16}
\begin{tikzpicture}   
\begin{axis}[
    legend columns=-1,
    legend entries={WSMP traffic, IP traffic, GPS tracking, Phishing attack, WSMP Flood, Geo-WSMP-Flood},
    legend to name=named,
    legend style={at={(0.5,-0.09)},
    anchor=north,legend columns=-1},
    symbolic x coords={Recall,F1-Score},
    major tick length=0cm,
    xtick=data,
    ymin=90.0,
     enlarge x limits=0.5,
    enlarge y limits={upper,value=0.2},
    nodes near coords,
    ybar,
    every node near coord/.append style={rotate=90, anchor=west},
    bar width = 15pt,
]

\addplot 
coordinates {(Recall,97.70) (F1-Score,98.38)}; 

\addplot 		
coordinates {(Recall,99.71) (F1-Score,99.64)};

\addplot 		
coordinates {(Recall,100) (F1-Score,99.80)};

\addplot 		
coordinates {(Recall,100) (F1-Score,99.73)};

\addplot 		
coordinates {(Recall,98.70) (F1-Score,98.06)};

\addplot 		
coordinates {(Recall,91.66) (F1-Score,95.65)};

\end{axis}
\end{tikzpicture}% NO EMPTY LINE HERE!!!! 
\begin{tikzpicture}  
\begin{axis}[
    legend style={at={(0.5,-0.09)},
    anchor=north,legend columns=-1},
    symbolic x coords={FPR,FNR},
    major tick length=0cm,
    xtick=data,
    ymin=0.0,
    enlarge x limits=0.5,
    enlarge y limits={upper,value=0.2},
    nodes near coords,
    ybar,
    every node near coord/.append style={rotate=90, anchor=west},
    bar width = 15pt,
]

\addplot 
coordinates {(FPR,0.16) (FNR,2.29)}; 

\addplot 		
coordinates {(FPR,0.39) (FNR,0.28)};

\addplot 
coordinates {(FPR,0.08) (FNR,0.00)}; 

\addplot 		
coordinates {(FPR,0.07) (FNR,0.00)};

\addplot 
coordinates {(FPR,0.14) (FNR,1.29)}; 

\addplot 		
coordinates {(FPR,0.00) (FNR,8.30)};

\end{axis}
\end{tikzpicture}
\end{adjustbox}

\begin{adjustbox}{max width=0.8\textwidth}
\ref{named}
\end{adjustbox}
\captionsetup{justification=centering}
\caption{Network Traffic multiclass classification results. Decision Tree Algorithm: Left - Recall, F1-Score. Right - FPR and FNR}

\label{fig:MultiClassMetrics}
\end{figure*}
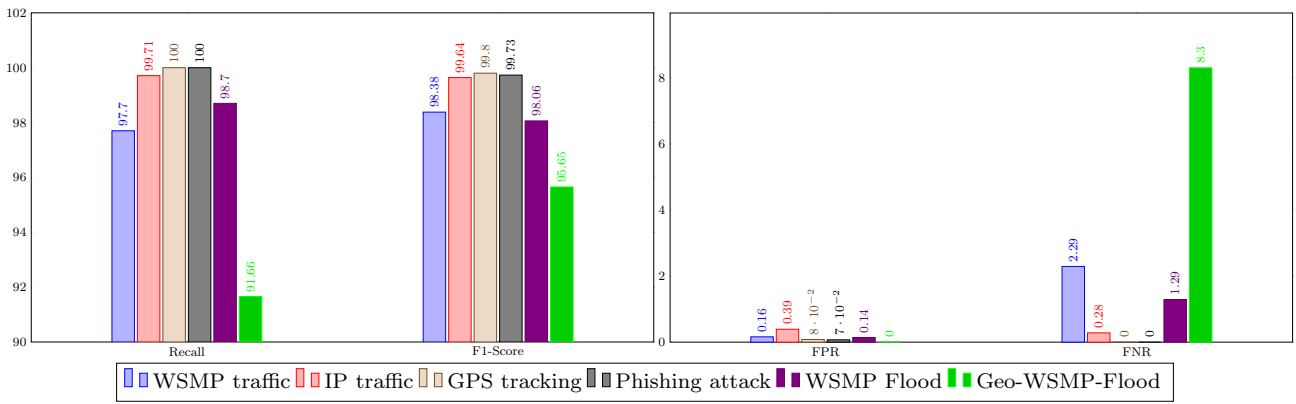
%%%%%%%%%%%%%%%%%%%%%%%%%%%%%%%%%%%%%%%%%%%%%%%%%%%%%%%%%%%%%%%%
Decision tree presents the best average recall, however, the malicious detection rate is found to be 79,70\%, which represents an intolerable false negative rate ($>$20 \%). Although the decision tree can recognize legitimate traffic with high precision, the malicious traffic detection performance is not promising. The poor detection rate of malicious traffic is due to the heterogeneity of legitimate traffic. Unlike the other types of networks, in vehicular networks, there are two types of network traffic: IP and WSMP. The two network traffic show different traffic patterns influenced by several factors such as the context of use, duration, size of packets, etc. This difference represents a challenge in training the classification model and subsequently mislead the classifier on differentiating between legitimate WSMP and malicious traffic. 
%=================== Important ===========================
% C'est un point très important qu'il faut developper un peu plus 
%=========================================================

To overcome the aforementioned issue, we create two classes of legitimate traffic by separating WSMP and IP traffic. We believe training the classification model with two separated categories of legitimate traffic, will reduce the false negative rate, and thus improve the malicious traffic detection rate. In addition, to provide a suitable response measure, the detection framework needs to identify the type of attacks. Therefore, we separate the malicious traffic into different classes: GPS tracking, WSMP flood, phishing, and Geo flood. From table \ref{tab:valiceRandomRes}, we can see a significant improvement in malicious traffic detection rate, the recall has increased from 79.7\% to 97.59\%. Figure \ref{fig:MultiClassMetrics} compare the recall, F1-score, FNR, and FPR for each traffic class separately. We see that decision tree achieves the best performances. Although, the slight decrease, the recall of the benign class remains high ($>$98). The detection rate of  Geo flood attack is not as good as the other attacks. This can be explained by the fact that the majority of security applications on the WSMP protocol use broadcast, which is quite similar to Geo flood attack.

According to the aforementioned results, it is important to note that the Geo WSMP flood attack represents the most challenging attack to detect. The Decision Tree achieved the highest detection rate with the lowest false positive rate. Besides detecting malicious traffic, it also delivers high performances on identifying the type of benign (WSMP, IP) and malicious traffic (GPS tracking, phishing, WSMP flood, Geo flood). Therefore, the decision tree is the best classifier to be selected for the network analyzer module.
%%%%%%%%%%%%%%%%%%%%%%%%%%%%%%%%%%%%%%%%%%%%%%%%%%%%%%%%%%%%%%%%

%%%%%%%%%%%%%%%%%%%%%%%%%%%%%%%%%%%%%%%%%%%%%%%%%%%%%%%%%%%%%%%%
\begin{figure*}[h!]
\centering
\begin{adjustbox}{width=0.98\textwidth}
\pgfplotsset{width=0.93\textwidth,height=0.4\textheight,compat=1.16}
\begin{tikzpicture}   
\begin{axis}[
    legend columns=-1,
    legend entries={KNN, Neural Networks, Decision Tree, Random Forest, SVM, Naive Bayes},
    legend to name=named,
    legend style={at={(0.5,-0.09)},
    anchor=north,legend columns=-1},
    symbolic x coords={Accuracy, F1-score},
    major tick length=0cm,
    xtick=data,
    %x tick label style={rotate=30,anchor=east},
    ymin=55.0,
     enlarge x limits=0.6,
    enlarge y limits={upper,value=0.2},
    nodes near coords,
    ybar,
    every node near coord/.append style={rotate=90, anchor=west},
    bar width = 15pt,
]

\addplot 
coordinates {(Accuracy,91.40) (F1-score,80.23)}; 

\addplot 		
coordinates {(Accuracy,83.60) (F1-score,56.70)}; 

\addplot 		
coordinates {(Accuracy,100) (F1-score,100)}; 

\addplot 		
coordinates {(Accuracy,99.9) (F1-score,99.90)}; 

\addplot 		
coordinates {(Accuracy,83.5) (F1-score,61.68)}; 

\addplot 		
coordinates {(Accuracy,99.9) (F1-score,99.83)}; 

\end{axis}
\end{tikzpicture}% NO EMPTY LINE HERE!!!! 
\begin{tikzpicture}  
\begin{axis}[
    legend style={at={(0.5,-0.09)},
    anchor=north,legend columns=-1},
    symbolic x coords={FPR, FNR},
    major tick length=0cm,
    xtick=data,
    %x tick label style={rotate=30,anchor=east},
    ymin=0.0,
    enlarge x limits=0.6,
    enlarge y limits={upper,value=0.2},
    nodes near coords,
    ybar,
    every node near coord/.append style={rotate=90, anchor=west},
    bar width = 15pt,
]

\addplot 
coordinates {(FPR,1.15) (FNR,19.66)}; 

\addplot 		
coordinates {(FPR,2.15) (FNR,37.55)};

\addplot 
coordinates {(FPR,0) (FNR,0.0)}; 

\addplot 		
coordinates {(FPR,0.0) (FNR,0.08)};

\addplot 
coordinates {(FPR,2.16) (FNR,37.79)}; 

\addplot 		
coordinates {(FPR,0) (FNR,0.16)};

\end{axis}
\end{tikzpicture}
\end{adjustbox}
\begin{adjustbox}{max width=0.8\textwidth}
\ref{named}
\end{adjustbox}
\captionsetup{justification=centering}
\caption{In-Vehicle Binary classification results: Left:Accuracy and F1-score. Right: FPR and FNR}
\label{fig:InVAccFPR}
\end{figure*}
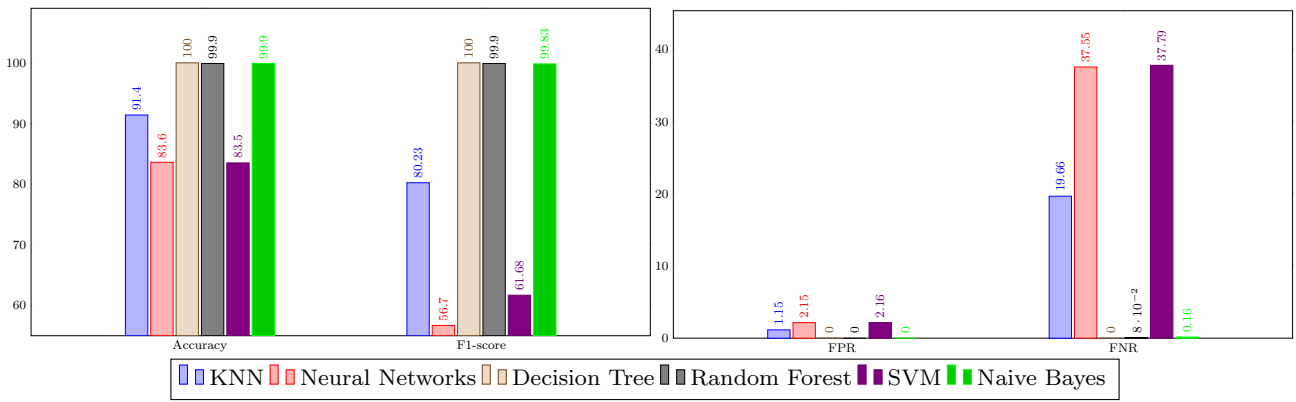
%%%%%%%%%%%%%%%%%%%%%%%%%%%%%%%%%%%%%%%%%%%%%%%%%%%%%%%%%%%%%%%%
\subsubsection{Malicious in-vehicle traffic detection}
We evaluate the performance of AntibotV on detecting and identifying in-vehicle attacks that can be carried out by bot malware. First, we carry out a binary classification (benign/malicious) using the supervised machine learning algorithms described earlier in section \ref{sec:analyzer}, then the best classification algorithm is assessed on identifying the type of attack. The accuracy bar charts of the figure  \ref{fig:InVAccFPR} clearly show that the proposed framework gives high performances on detecting and identifying the in-vehicle traffic, with over 90\% for the KNN, decision tree, random forest, and naive Bayes algorithms. The results of binary classification are presented in table \ref{tab:InVBinRes}. The decision tree is found to deliver the maximum detection rate for both benign and malicious frames. KNN, Neural Networks and SVM algorithms failed to detect the class of malicious traffic, with a poor detection rate (between 25\% and 61\%). From table \ref{tab:valiceDecisionMulti}, we can see that the decision tree can perfectly identify the different types of in-vehicle attacks. These results confirm the efficiency of tree-based algorithms for intrusion detection. Therefore, the decision tree is the best classifier to be selected for the in-vehicle analyzer module.
%%%%%%%%%%%%%%%%%%%%%%%%%%%%%%%%%%%%%%%%%%%%%%%%%%%%%%%%%%%%%%%%
\begin{table*}[h!]
  \caption{In-Vehicle traffic binary classification using AntibotV}
  \label{tab:InVBinRes}
\begin{adjustbox}{width=0.8\textwidth, center}
\begin{tabular}{cccccccc}
\hline
Algorithm                          & Classe           & Precision & Recall   & F1-score & FPR    & FNR     & Accuracy                  \\ \hline
\multirow{3}{*}{KNN}               & Benign frames    & 99,90\%   & 100,00\% & 99,90\%  & 0,10\% & 0,00\%  & \multirow{3}{*}{91,40\%}  \\
                                   & Malicious frames & 61,30\%   & 60,58\%  & 60,55\%  & 2,20\% & 39,33\% &                           \\
                                   & Average value    & 80,60\%   & 80,29\%  & 80,23\%  & 1,15\% & 19,66\% &                           \\ \hline
\multirow{3}{*}{Neural   Networks} & Benign frames    & 100,00\%  & 100,00\% & 100,00\% & 0,00\% & 0,00\%  & \multirow{3}{*}{83,60\%}  \\
                                   & Malicious frames & 12,48\%   & 24,85\%  & 13,40\%  & 4,30\% & 75,10\% &                           \\
                                   & Average value    & 56,24\%   & 62,43\%  & 56,70\%  & 2,15\% & 37,55\% &                           \\ \hline
\multirow{3}{*}{Decision   Tree}   & Benign frames    & 100,00\%  & 100,00\% & 100,00\% & 0,00\% & 0,00\%  & \multirow{3}{*}{100,00\%} \\
                                   & Malicious frames & 100,00\%  & 100,00\% & 100,00\% & 0,00\% & 0,00\%  &                           \\
                                   & Average value    & 100,00\%  & 100,00\% & 100,00\% & 0,00\% & 0,00\%  &                           \\ \hline
\multirow{3}{*}{Random   Forest}   & Benign frames    & 100,00\%  & 100,00\% & 100,00\% & 0,00\% & 0,00\%  & \multirow{3}{*}{99,90\%}  \\
                                   & Malicious frames & 99,83\%   & 99,83\%  & 99,80\%  & 0,01\% & 0,15\%  &                           \\
                                   & Average value    & 99,91\%   & 99,91\%  & 99,90\%  & 0,00\% & 0,08\%  &                           \\ \hline
\multirow{3}{*}{SVM}               & Benign frames    & 100,00\%  & 100,00\% & 100,00\% & 0,00\% & 0,00\%  & \multirow{3}{*}{83,50\%}  \\
                                   & Malicious frames & 23,33\%   & 24,33\%  & 23,35\%  & 4,33\% & 75,58\% &                           \\
                                   & Average value    & 61,66\%   & 62,16\%  & 61,68\%  & 2,16\% & 37,79\% &                           \\ \hline
\multirow{3}{*}{Naive   Bayes}     & Benign frames    & 100,00\%  & 100,00\% & 100,00\% & 0,00\% & 0,00\%  & \multirow{3}{*}{99,90\%}  \\
                                   & Malicious frames & 99,68\%   & 99,65\%  & 99,65\%  & 0,00\% & 0,33\%  &                           \\
                                   & Average value    & 99,84\%   & 99,83\%  & 99,83\%  & 0,00\% & 0,16\%  &                           \\ \hline
\end{tabular}
\end{adjustbox}
\end{table*}
%%%%%%%%%%%%%%%%%%%%%%%%%%%%%%%%%%%%%%%%%%%%%%%%%%%%%%%%%%%%%%%
\begin{table*}[h!]
  \caption{In-Vehicle multiclass results: AntibotV based on Decision Tree classifier}
  \label{tab:valiceDecisionMulti}
\small
\begin{adjustbox}{width=0.8\textwidth, center}
\begin{tabular}{cccccccc}
\hline
\multicolumn{2}{c}{Classes}                        & Precision & Recall   & F1-score & FPR    & FNR    & Accuracy                  \\ \hline
\multicolumn{2}{c}{Benign frames}                  & 100,00\%  & 100,00\% & 100,00\% & 0,00\% & 0,00\% & \multirow{6}{*}{100,00\%} \\ \cline{1-7}
\multirow{4}{*}{Malicious   frames} & DoS Attack   & 100,00\%  & 100,00\% & 100,00\% & 0,00\% & 0,00\% &                           \\
                                    & Fuzzy Attack & 100,00\%  & 100,00\% & 100,00\% & 0,00\% & 0,00\% &                           \\
                                    & Gear Attack  & 100,00\%  & 100,00\% & 100,00\% & 0,00\% & 0,00\% &                           \\
                                    & RPM Attack   & 100,00\%  & 100,00\% & 100,00\% & 0,00\% & 0,00\% &                           \\ \cline{1-7}
\multicolumn{2}{c}{Average Value}                  & 100,00\%  & 100,00\% & 100,00\% & 0,00\% & 0,00\% &                           \\ \hline
\end{tabular}
\end{adjustbox}
\end{table*}

%%%%%%%%%%%%%%%%%%%%%%%%%%%%%%%%%%%%%%%%%%%%%%%%%%%%%%%%%%%%%%%%
\subsection{Discussion and comparison}
We compared our solution against the solution presented in \cite{garip2016ghost} that tackled the detection of vehicular botnets, the authors used anomaly detection technique to allow detection of new forms of injected BSMs messages. The detection solution assumes a specific botnet communication protocol (GHOST), which makes it unable to detect botnet using a different communication protocol. The proposed framework AntibotV does not suppose a particular botnet communication protocol, so the detection approach works independently of the botnet communication protocol. Limiting detection to monitoring only network communication, makes \cite{garip2016ghost} unable to detect in-vehicle attacks. Thanks to two-level monitoring, AntibotV can also detect in-vehicle attacks. Compared to \cite{garip2016ghost}, AntibotV delivers better accuracy and lower false positive rate (see table \ref{tab:Comp}). In \cite{seo2018gids}, the authors considered in-vehicle threats, they proposed an anomaly detection framework based on convolutional neural network and deep neural network. Although AntibotV delivers better performances, the proposed approach \cite{seo2018gids} can detect unseen attacks, but to the detriment of false alerts. Furthermore, deep learning techniques require a large amount of computational and memory resources, which could be constraining in vehicular context. The authors in \cite{agrawal2021sdn} have proposed a mechanism which, detects, mitigates and traces back the location of a Low-Rate DDoS attacker. If the DDoS attack is caused by a botnet, the traceability mechanism can be used to identify bot nodes in the network. However, if the botnet is used to apply other threats (e.g. information theft), the proposed detection techniques will not be effective anymore, unlike our proposed framework (Antibotv), which is able to deal with different types of botnet threats at different levels. Table \ref{tab:Comp} provides a qualitative comparison between AntibotV and \cite{garip2016ghost, agrawal2021sdn} (different databases), and a quantitative comparison with \cite{seo2018gids} (the same in-vehicular traffic database).

%Table \ref{tab:Comp} provides a comparison between AntibotV and existing solutions
%%%%%%%%%%%%%%%%%%%%%%%%%%%%%%%%%%%%%%%%%%%%%%%%%%%%%%%%%%%%%%%%%

\begin{table*}[h!]
  \caption{Comparison among vehicular botnets detection systems}
  \label{tab:Comp}
\small
\begin{adjustbox}{width=1.0\textwidth,center}
\begin{tabular}{ccccc}
\hline
                           & AntibotV (our work)                                                                                                                             &\cite{garip2016ghost} 
                           & \cite{seo2018gids} 
                           & \cite{agrawal2021sdn} \\ \hline
Threat level               & Network   \& In-vehicle                                                                                                                         & Network                                & In-vehicle                          & Network                              \\
Machine learning technique & Classification                                                                                                                                  & Anomaly   detection                    & Anomaly   detection                 & Time-series analysis                       \\
Algorithm                  & Decision   Tree                                                                                                                                 & 3D DBSCAN                              & CNN   \& DNN                        & XGBoost                   \\
Dataset                    & \begin{tabular}[c]{@{}c@{}}Our own network traffic dataset,\\ and \cite{seo2018gids} in-vehicle traffic dataset\end{tabular} & Their own dataset                      & Their own dataset                   & Their own dataset                    \\
Ressources requirement     & Low                                                                                                                                             & Low                                    & High                                & Low                                  \\
Accuracy                   & 99.40\% \& 100\%                                                                                                                                & 77\%                                   & 97.53\%                             & 97.6\%                                 \\
Detection rate (Recall)    & 97.96\% \& 100\%                                                                                                                                & NA                                     & 98.65\%                             & NA                              \\
Precision                  & 99.19\% \& 100\%                                                                                                                                & NA                                     & 97.63\%                             & NA                              \\
FPR                        & 0.14\% \& 0\%                                                                                                                                   & NA                                     & NA                                  & 3.29\%                                   \\ \hline
\end{tabular}
\end{adjustbox}
\end{table*}

%%%%%%%%%%%%%%%%%%%%%%%%%%%%%%%%%%%%%%%%%%%%%%%%%%%%%%%%%%%%%%%%%%

\section{Conclusion} \label{sec:6}
In this paper, we have proposed AntibotV, a multilevel behaviour-based framework to detect vehicular botnets. We have considered new zero-day attacks, as well as a wide range of DoS and in-vehicle attacks. The proposed framework monitors the vehicle’s activity at network and in-vehicle levels. To build the detection system, we have collected network traffic data of legitimate and malicious applications. Then, training using decision tree a new classifier with a set of features that we have extracted and selected. Likewise, we have trained a decision tree with in-vehicle data. The experimental results showed that AntibotV outperforms existing solutions, it achieves a detection rate higher than 97\% and a false positive rate lower than 0.14\%. To add a realistic dimension to our framework, the ideal was to do the training and validation with a dataset generated from a realistic testbed. In our future work, we will work on the generation of a realistic dataset, which takes into account different factors, such as the impact of the environment (urban, rural and highway), plus the implementation of different malicious behaviour of a bot vehicle. In a second step, we can use AntibotV generated from real data to design a simple system to report feedback from the driver, instead of using only raw features directly. Moreover, AntibotV has a fundamental limitation in detecting unlearned types of attacks because it is based on supervised learning. To solve this problem, further research on detecting unknown attacks and the updating process is needed using advanced learning techniques. We intend to label legitimate traffic using anomaly detection techniques and categorizing malicious traffic using an ensemble-based classifier.

\section{Acknowledgment}
This research is a result from PRFU project C00L07UN23 0120180009 funded in Algeria by La Direction Générale de la Recherche Scientifique et du Développement Technologique (DGRSDT).
%%%%%%%%%%%%%%%%%%%%%%%%%%%%%%%%%%%%%%%%%%%%%%
%%%%%%%%%%%%%%% Fin Test %%%%%%%%%%%%%%%%%%%%%%%%%
%%%%%%%%%%%%%%%%%%%%%%%%%%%%%%%%%%%%%%%%%%%%%%%%

%\newpage
\bibliographystyle{unsrt}
\bibliography{mybibfile.bib}   % name your BibTeX data base

%\newpage
\begin{wrapfigure}{l}{25mm} 
    \includegraphics[width=1in,height=1.25in,clip,keepaspectratio=true]{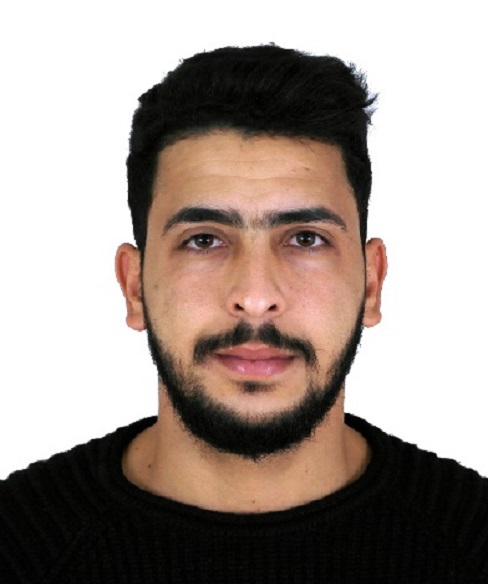}
  \end{wrapfigure}\par
\textbf{Rabah RAHAL} is a doctoral student in Badji Mokhtar University Annaba Algeria. He is a member of the Laboratory of Computer Networks and Systems. His research interests include vehicular networks, network security, intrusion and malwares detection, machine learning, anomaly detection,
cryptography, security analysis. He does research in cyberattacks detection in vehicular networks, and the generation of realistic DoS dataset for intelligent transportation systems.\par

\begin{wrapfigure}{l}{25mm} 
    \includegraphics[width=1in,height=1.25in,clip,keepaspectratio=true]{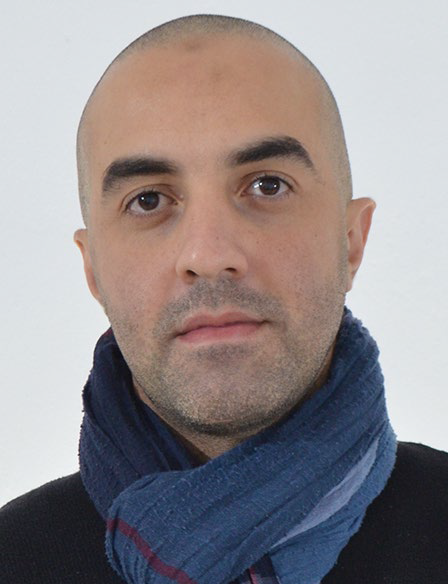}
  \end{wrapfigure}\par
  \textbf{Abdelaziz Amara korba} received his Ph.D. degree from Badji Mokhtar Annaba University, Algeria, in 2016. Currently, he is an Associate Professor in the department of computer science at the University of Badji Mokhtar Annaba, and he is member of Networks and Systems Laboratory (LRS). Dr. Amara korba has made contributions in the fields of network security, intrusion and cyberattacks detection. His research interests include cybersecurity, intrusion detection, anomaly detection, IoT, botnets. He served as TPC in many international conferences worldwide (GLOBECOM, EUSPN, AINIS, ICCE,..), he serves as an associate editor for the International Journal of Smart Security Technologies.\par
  
  \begin{wrapfigure}{l}{25mm} 
    \includegraphics[width=1in,height=1.25in,clip,keepaspectratio=true]{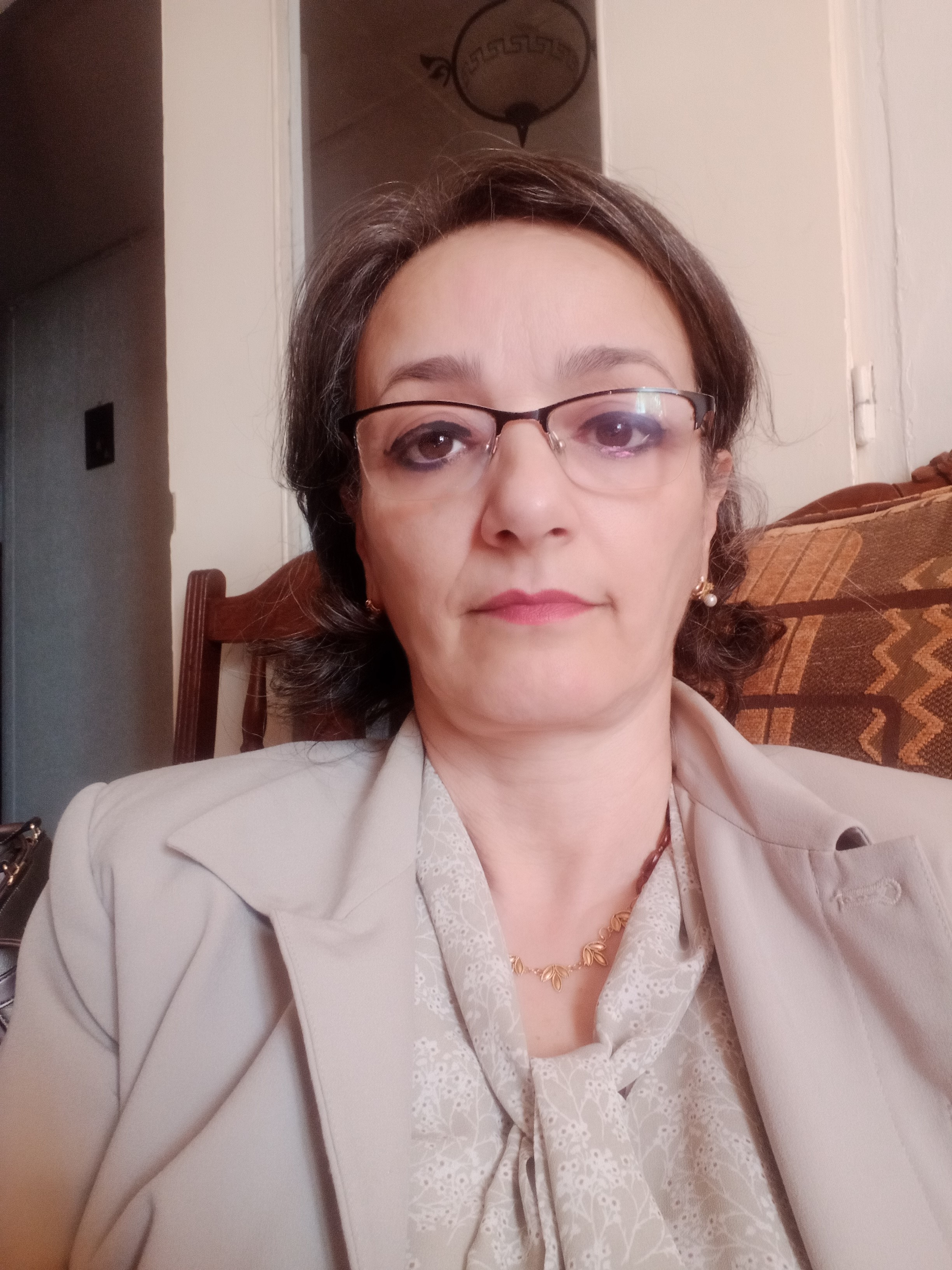}
  \end{wrapfigure}\par
  \textbf{Nacira Ghoualmi‑Zine} is a Professor in Computer Sciences and has been a Lecturer in the Department of Computer Science at Badji Mokhtar University, Annaba, Algeria since 1985. She is the Head of the Master and Doctoral option entitled Network and Computer Security, and Head of a Laboratory of Computer Networks and Systems. Her research includes cryptography, computer security, intrusion detection system, wireless networks, distributed multimedia applications, quality of service, security in the protocol, and optimisation in networks.\par
 
%\newpage
  \begin{wrapfigure}{l}{25mm} 
    \includegraphics[width=1in,height=1.25in,clip,keepaspectratio=true]{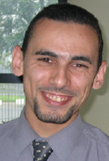}
  \end{wrapfigure}\par

\textbf{Yacine CHALLAL} received the PhD degree in computer science from Compigne University of Technology (UTC), France, in 2005. Since September 2007, he has been an Assistant Professor at UTC. He is currently Professor Ecole nationale Supérieure d’Informatique, Algiers, Algeria. His research interests include network security, wireless communication security, cyber-physical systems security, wireless sensor networks, IoT, Cloud computing and fault tolerance in distributed systems. He served as managing guest editor of Ad Hoc Networks for a special issue on Internet of Things security and privacy, and served as guest editor of Wiley Security and Communication Networks journal on Security, Privacy and Trust in Emerging Wireless Networks.\par

\begin{wrapfigure}{l}{25mm} 
    \includegraphics[width=1in,height=1.25in,clip,keepaspectratio=true]{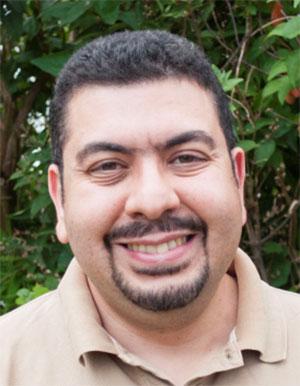}
  \end{wrapfigure}\par
\textbf{Yacine Ghamri-Doudane} is currently Full Professor at the University of La Rochelle (ULR) in France, and the director of its Laboratory of Informatics, Image and Interaction (L3i). Before that, Yacine held an Assistant/Associate Professor position at ENSIIE, a major French post-graduate school located in Evry, France, and was a member of the Gaspard-Monge Computer Science Laboratory (LIGM – UMR 8049) at Marne-la-Vallée, France. Yacine received an engineering degree in computer science from the National Institute of Computer Science (INI), Algiers, Algeria, in 1998, an M.S. degree in signal, image and speech processing from the National Institute of Applied Sciences (INSA), Lyon, France, in 1999, a Ph.D. degree in computer networks from University Pierre \& Marie Curie, Paris 6, France, in 2003, and a Habilitation to Supervise Research (HDR) in Computer Science from Université Paris-Est, in 2010. His current research interests lays in the area of wireless networking and mobile computing with a current emphasis on topics related to the Internet of Things (IoT), Wireless Sensor Networks (WSN), Vehicular Networks as well as Digital Trust. Yacine holds three (3) international patents and he authored or co-authored seven (7) book chapters, more than 30 peer-reviewed international journal articles and more than 130 peer-reviewed conference papers.\par

\end{document}